\documentclass[11pt,a4paper]{article}

 \usepackage[latin1]{inputenc}
 \usepackage[english]{babel}
 \usepackage{myway}

\hypersetup{pdftitle={D7-Inflation}, 
pdfauthor={Arthur Hebecker, Sebastian Kraus, Dieter L{\"u}st, Stephan 
Steinfurt, Timo Weigand}, 
pdfsubject={String theory model building},
bookmarksopen, bookmarksnumbered, bookmarksopenlevel=2}

\newcommand{\beq}{\begin{equation}} 
 \newcommand{\eeq}{\end{equation}}
\newcommand{\bal}{\begin{aligned}}  
 \newcommand{\eal}{\end{aligned}}
\newcommand{\bea}{\begin{eqnarray}} 
 \newcommand{\eea}{\end{eqnarray}}
 \def\ov{\overline}

\newcommand{\bbF}{\mathbb{F}}
\newcommand{\ccA}{\mathcal{A}}

\newcommand{\ccF}{\mathcal{F}}
\newcommand{\ccB}{\mathcal{B}}

\newcommand{\myfrac}[2]{\genfrac{[}{]}{0pt}{}{#1}{#2}}

\title{${\mathbb F}$luxbrane Inflation}

\preprint{MPP-2011-44\\
LMU-ASC 14/11\\
CERN-PH-TH/2011-088}

\author[1]{Arthur Hebecker\note[ ]{\href{mailto:A.Hebecker@ThPhys.Uni-Heidelberg.de}{A.Hebecker@ThPhys.Uni-Heidelberg.de}},}
\author[1]{Sebastian C. Kraus\note[ ]{\href{mailto:S.Kraus@ThPhys.Uni-Heidelberg.de}{S.Kraus@ThPhys.Uni-Heidelberg.de}},}
\author[2]{Dieter L{\"u}st\note[ ]{\href{mailto:Dieter.Luest@lmu.de}{Dieter.Luest@lmu.de}},}
\author[1]{Stephan Steinfurt\note[ ]{\href{mailto:S.Steinfurt@ThPhys.Uni-Heidelberg.de}{S.Steinfurt@ThPhys.Uni-Heidelberg.de}},}
\author[1]{\\ and Timo Weigand\note[ ]{\href{mailto:T.Weigand@ThPhys.Uni-Heidelberg.de}{T.Weigand@ThPhys.Uni-Heidelberg.de}}}
\affiliation[1]{Institut f\"ur Theoretische Physik, Universit\"at Heidelberg, Philosophenweg 19, D-69120 Heidelberg\vspace{0.1cm}}
\affiliation[2]{Arnold-Sommerfeld-Center, Ludwig-Maximilians-Universit\"at, Theresienstrasse 33, \\ \phantom{a} D-80333 M{\"u}nchen\vspace{0.1cm}}
\affiliation[2]{Max-Planck-Institut f\"ur Physik, F\"ohringer Ring 6, D-80805 M\"unchen\vspace{0.1cm}}
\affiliation[2]{Physics Division (TH), CERN, 1211 Geneva 23, Switzerland}

\abstract{ 
As a first step towards inflation in genuinely ${\mathbb F}$-theoretic 
setups, we propose a scenario where the inflaton is the relative position 
of two 7-branes on holomorphic 4-cycles. Non-supersymmetric gauge flux 
induces an attractive inter-brane potential. The latter is sufficiently 
flat in the supergravity regime of large volume moduli. Thus, in contrast to 
brane-antibrane inflation, fluxbrane inflation does not require warping. 
We calculate the inflaton potential both in the supergravity approximation 
and via an open-string one-loop computation on toroidal backgrounds. This 
leads us to propose a generalisation to genuine Calabi-Yau manifolds. We 
also comment on competing $F$-term effects. The end of inflation is marked 
by the condensation of tachyonic recombination fields between the 7-branes, 
triggering the formation of a bound state described as a stable extension 
along the 7-brane divisor. Hence our model fits in the framework of hybrid 
$D$-term inflation. We work out the main phenomenological properties of 
our $D$-term inflaton potential. In particular, our scenario of D7/D7 
inflation avoids the familiar observational constraints associated with 
cosmic strings. 
}

\begin{document}

\maketitle

\section{Introduction}

In its pursuit of a unified picture of particle physics and cosmology, string phenomenology
has found a particularly fruitful arena in the framework of Type II compactifications with branes.
On the one hand, the open string sector can in principle encompass the essential features of the Standard Model. On the other hand, a quantitative treatment of cosmology is within reach thanks to an improved understanding of moduli stabilisation. The latter is, for instance, a prerequisite for a reliable computation of the dynamics of candidate inflaton fields.

In particular, Type IIB Calabi-Yau orientifolds with branes and fluxes and their strong coupling version of F/M-theory on Calabi-Yau fourfolds are  a corner in the string landscape
where important progress in \emph{both} directions  - particle physics and scalar field dynamics - has been made.
For example, on the gauge theory side it has recently been appreciated that F-theory \cite{Vafa:1996xn}
is a promising framework for constructing realistic GUT string vacua
\cite{Beasley:2008dc,Donagi:2008ca} (see \cite{Heckman:2010bq,
Weigand:2010wm} for recent reviews). 
At the same time  a combination of  $G_4$ fluxes and non-perturbative effects can in principle achieve moduli stabilisation within the framework of conformal Calabi-Yau and in particular K\"ahler compactifications. This is essential for concrete model building and hands-on 
applications since our understanding of non-K\"ahler geometries is at 
present still limited. 

Compared to the recent developments in particle physics model building, there is far less progress 
concerning the implementation of inflation in F-theory, preferably in the 
same class of  models with already attractive particle phenomenology.
One possibility would be to pursue the embedding of existing inflationary scenarios such as D3/(anti)D3 or D3/D7-brane inflation into F-theory models.
On the other hand, genuine and possibly non-perturbative F-theoretic effects are 
expected, in particular, in the 7-brane sector. 
An obvious first step in 
this direction is to study  inflationary D7-brane dynamics. In this spirit, the 
present paper embarks on the analysis of perturbative D7/D7
inflation in Type IIB orientifolds. This turns out to be in itself an 
extremely interesting setting with attractive phenomenological features. 

As a motivation for our scenario of fluxbrane inflation, we recall that the appealing idea 
of brane-antibrane inflation \cite{Dvali:1998pa} faces a well-known 
phenomenological problem: Given the limited range available for brane 
positions inside the compact geometry, the brane-antibrane potential is not 
sufficiently flat \cite{Burgess:2001fx}. 
Several approaches to circumvent this problem have been suggested. They 
include replacing the brane-antibrane pair by a pair of branes 
at a relative angle \cite{GarciaBellido:2001ky}, by a 
D7-brane with flux together with a D3-brane \cite{Dasgupta:2002ew}, 
and exploiting warped geometries \cite{Kachru:2003sx}. A closely related 
idea is that of Wilson line inflation \cite{Avgoustidis:2006zp} (also known 
in a field-theoretic context \cite{ArkaniHamed:2003wu,Kaplan:2003aj}). 

In the present paper, we assume that inflation is driven by the 
relative motion of two spacetime-filling 7-branes. The inflationary 
potential results from a SUSY-breaking gauge flux on a 2-cycle shared 
by these two branes. The dominant part of this potential is a constant
$\sim |\ccF|^2$, where $\ccF$ is the 2-form flux on the branes.\footnote{Here $|\ccF|^2 = \ccF_{MN} \ccF_{PQ} g^{MP} g^{NQ}$ and we assume the use of coordinates which make $g_{MN}$ locally approximately Euclidean. Since $\ccF$ is integrally quantised, $\int \ccF = p \in \mathbb{Z}$, this implies $|\ccF|^2 \sim p^2 / R^4$, with $R$ a generic Calabi-Yau radius in units of $\ell_s$.} Reheating 
occurs when the two branes nearly coincide so that a tachyon develops and
a bound state forms. This is clearly consistent with a leading-order 
analysis of $D$-term hybrid inflation \cite{Binetruy:1996xj,Halyo:1996pp}, 
which indeed turns out to be the correct 4d supergravity 
description.\footnote{
Note 
that $F$-terms induced by background or gauge fluxes can stabilise the 7-brane 
modulus at a higher scale; only those 7-branes can lead to inflation for 
which this $F$-term effect is absent or sufficiently small due to a suitable 
choice of fluxes. This constraint will be analysed further in the corpus of 
this paper.}
A brane-to-brane force, i.e.\ a non-constant contribution to the energy, arises 
as a Coleman-Weinberg-type loop correction to the $D$-term potential. We 
derive this effect both from a string-loop calculation and in 10d supergravity.
In the latter approach, it can be seen as a classical force between a flux 
$\ccF/2$ on one of the branes and the opposite flux $-\ccF/2$ on its 
partner. In contrast to the $|\ccF|^2$ scaling of the constant term, this 
Coulomb-like force scales as $|\ccF|^4$. We will give an intuitive argument for 
this crucial feature of our potential in a moment. It is this particular 
type of scaling which allows for a sufficiently flat potential in the 
weak-flux (i.e.\ large volume) limit. 

While our scenario is conceptually similar to D3/D7 inflation \cite{
Dasgupta:2002ew,Hsu:2003cy,Haack:2008yb}, it has significant advantages 
concerning the field range of the inflaton: The fact that a D7- rather than a 
D3-brane moves in the compact space enhances the relevant kinetic term. This, 
in turn, gives the canonically normalised inflaton field a much larger field 
range (see, however, the `anisotropic compact space' proposal in \cite{
Haack:2008yb} and the possibility of having the D7-brane move in the 
background of a large-$N$ D3-brane stack \cite{Hsu:2003cy}). 
As a second considerable virtue, we will find that our scenario passes the cosmic string constraints more easily than 
generic models of $D$-term inflation.

Like many other brane inflation models, D7/D7 inflation 
requires a certain amount of fine tuning after K\"ahler moduli stabilisation. 
However, the mechanisms available for this tuning are rather special. As 
it turns out, truly F-theoretic effects come into play in this context.

\subsubsection*{Summary of technical results}

After this general overview we would now like to enter a slightly more detailed and technical discussion of the issues just described.
We first recall the familiar `no-go theorem' of \cite{Burgess:2001fx}: 
Schematically, the 4d Lagrangian for the relative motion of a $p$-brane 
and its anti-brane in the compactification space reads
\beq
{\cal L} \quad \sim \quad g_s^{-2}\,{\cal V}\,\,{\cal R} \,\,\, + \,\,\, 
g_s^{-1}\,{\cal V}_{\scriptscriptstyle ||}\,\left[\,(\partial r)^2\, - \, 
\left(\,A-B\,\frac{g_s}{r^{d_\perp-2}}\,\right)\,\right]\,.
\label{lag}
\eeq
Here $g_s$ is the string coupling and ${\cal R}$ is the 4d Ricci scalar. 
All 10d quantities, such as the total and brane-parallel compactification 
volumes ${\cal V}$ and ${\cal V}_{\scriptscriptstyle ||}$ as well as the 
brane-antibrane separation $r$, are measured in units of the string length. 
Furthermore, we have chosen our 4d coordinates and hence ${\cal L}$ to be 
dimensionless. Finally, $d_\perp=9-p$ is the codimension of the $p$-branes 
and $A,B$ are positive ${\cal O}(1)$ constants. 

The potential of our inflaton $r$ in (\ref{lag}) consists of two pieces: 
A constant part (\mbox{$\sim A$}) associated with the brane tensions and a 
Coulomb-like attractive contribution ($\sim B$). After reheating, which 
corresponds to brane-anti-brane annihilation and is outside the validity 
range of (\ref{lag}), the potential is identically zero. 

It is now an easy exercise to go to the Einstein frame, to normalize the 
inflaton canonically, and to calculate the slow roll parameter 
$\eta\equiv V''/V$. Introducing the brane-perpendicular size $L_\perp$ 
of the compact space, ${\cal V}={\cal V}_{\scriptscriptstyle ||}L_\perp^{d_\perp}$, 
the result reads
\beq
-\eta\,\,\,\sim\,\,\,\frac{B}{A}\,\left(\frac{L_\perp}{r}\right)^{d_\perp}\,.
\label{eta}
\eeq
It is immediately clear that $-\eta\ll 1$ cannot be realised since the 
brane-separation $r$ is bounded by $L_\perp$. This result also holds for 
$d_\perp=2$, which is in fact our particular focus. In this case, 
(\ref{lag}) has to be interpreted according to
\beq
\frac{1}{r^{d_\perp-2}} \,\to\, \ln(1/r)\qquad\mbox{for}\qquad d_\perp=2\,,
\eeq
with the rest of the calculation leading to (\ref{eta}) unchanged.

The suggestion of the present paper is to consider two D$p$-branes which we 
assume to carry 2-form flux $\ccF/2$ and $-\ccF/2$. Thinking of $\pm\ccF/2$ 
as of $(p-2)$-brane charges dissolved within each of the $p$-branes 
\cite{Douglas:1995bn}, one expects an attractive force and hence brane motion 
(inflation). When the branes finally come to be on top of each other, the 
flux annihilates and supersymmetry is restored (reheating). Given this 
intuitive picture, one might expect that (\ref{lag}) remains valid, but with 
both $A$ and $B$ now being $\sim |\ccF|^2$. In that case $\ccF$ would drop out 
of the formula for $\eta$ and the previous negative conclusion would still 
follow. Instead, as we will demonstrate, one finds the parametrical behaviour
\beq
A\sim |\ccF|^2\qquad\mbox{and}\qquad B\sim  |\ccF|^4\, .
\eeq
Thus, 
\beq
-\eta\,\,\,\sim\,\,\, |\ccF|^2\left(\frac{L_\perp}{r}\right)^{d_\perp}\,\,\ll
\,\,1
\eeq
in the limit of weak flux. The latter is easily realised 
by choosing an ${\cal O}(1)$ number of flux quanta $p$ and going to the limit of 
large brane volume, using $|\ccF|^2 \sim p^2 / R^4$. Hence, at least before moduli stabilisation, `fluxbrane 
inflation' arises naturally, without a fine-tuned cancellation 
between several competing contributions to the potential. 

Obviously, the crucial point is the absence of a force $\sim |\ccF|^2$, which 
makes the next term in a small-$|\ccF|^2$ expansion dominant and leads to the 
scaling $B\sim  |\ccF|^4$. We will now try to provide an intuitive 
understanding of this behaviour: We first recall that, due to the 
gauge-non-invariance of $B_2$ and its interplay with the DBI action, 
brane-flux is a {\it relative} concept. In other words, two branes with 
flux $\ccF/2$ and $-\ccF/2$ can be equivalently thought of as a brane with 
flux $\ccF$ in the background of a fluxless brane. Let us now focus on 
D3-branes (although this is clearly not the phenomenologically interesting 
case) and, in addition, replace the fluxless brane by a stack of $N$ 
D3-branes. Thus, the brane with flux $\ccF$ moves in the familiar 
AdS$_5\times$S$^5$ background \cite{Maldacena:1997re,Aharony:1999ti} and the 
only effect of this motion on its action is through a rescaling of the 
induced metric. However, due to the classical scale-invariance of the 
leading-order 4d gauge theory Lagrangian, the energy-density associated with 
the flux $\ccF$ is insensitive to this rescaling. In other words, there is no 
force $\sim |\ccF|^2$. Starting from this observation, it is easy to convince 
oneself that this situation persists for $p\neq 3$: While the gauge theory 
Lagrangian is not any more scale invariant in $d\neq 4$, the corresponding 
non-trivial $r$ dependence is exactly compensated by the non-trivial dilaton 
background that is sourced by D$p$-brane stacks with $p\neq 3$. As a result, 
the force remains $\sim |\ccF|^4$ for general fluxbrane-pairs. In fact, this is 
closely related to the {\it Generalised Conformal Symmetry} of ${p+1}$ 
dimensional Yang-Mills theory and the corresponding supergravity background 
discussed in \cite{Jevicki:1998yr,Jevicki:1998qs,Jevicki:1998ub}.\footnote{
We 
only require the classical version of this symmetry introduced in
Sect.\ 2.2 of \cite{Jevicki:1998ub}.}

Having now explained the basic idea in an intuitive way, we jump ahead and 
briefly present a more technical supergravity formulation of our results. 
The leading term in the potential, i.e.\ the Einstein-frame version of the 
term $\sim A$ from (\ref{lag}), can be interpreted as a $D$-term potential 
with field-dependent FI-term~$\xi$:
\beq
V=\frac{g_{\rm YM}^2\xi^2}{2}\qquad\mbox{and}\qquad \xi\sim \frac{g_s}{\cal V}\int J
\wedge {\cal F}\,.\label{pot0}
\eeq
Here ${\cal V}\sim\int J^3$ measures the Calabi-Yau volume in units of the 
string length and the field strength is now normalised as an integral 2-form.

The correction $\sim B$ with its logarithmic inflaton dependence 
was previously argued to arise from a Coulomb-like force between fluxbranes. 
In the case of parallel branes on a torus, we will provide a detailed
derivation both in 10d supergravity (corresponding to the Coulomb-force 
point of view) and through a string-loop calculation (corresponding to a 
Coleman-Weinberg-type correction in 4d field theory). This stringy derivation
turns out to be much more than just a consistency-check: As we will see, 
the moment of 60 e-foldings before reheating corresponds to a brane 
separation far below the string length. Thus, we cannot a priori 
trust the 10d-supergravity-derived potential when calculating CMB fluctuations. 
However, the stringy calculation demonstrates the validity of this potential 
up to the very point of tachyon condensation. 

The results outlined above will allow us to make a proposal for the generic 
Calabi-Yau situation, where the branes are only locally (approximately) 
parallel and flat. The corresponding potential reads\footnote{
Note 
that, in contrast to (\ref{lag}), this expression is in the 4d Einstein frame.
}
\beq
V=\frac{g_{\rm YM}^2\xi^2}{2}\left\{1+\frac{g_{\rm YM}^2}{16\pi^2}\left[\frac{1}{\left(\frac{1}{2}\int J^2\right)}\left(\int J\wedge{\cal F}\right)^2-4\left(\frac{1}{2}\int{\cal F}\wedge
{\cal F}\right)
\right]\,\ln(r)\right\}\,.\label{pot}
\eeq
Here we immediately recognize the terms $\sim A$ and $\sim B$ scaling with 
the second and fourth power of the flux.

The weak-flux or large-volume limit now arises in a slightly different 
manner: Before, we had $|\ccF|^2 \sim p^2/R^4$. Now, the $R$-dependence is encoded entirely in $J$. Thus, $J\sim R^2$ and $1/g_{\rm YM}^2\sim \int J^2 \sim R^4$ 
are responsible for the correct scaling and hence for the parametric 
smallness of the $\eta$-parameter following from (\ref{pot}). 

Another crucial point, which is apparent in (\ref{pot}) and which 
distinguishes our scenario from a stringy realisation of generic 
loop-corrected $D$-term inflation, is the following: Let us choose a flux 
which does not induce a D3 charge, i.e.\ a flux satisfying $\int{\cal F}^2
=0$. Then the $\ln(r)$ piece of the potential is suppressed by 
$(\int J\wedge {\cal F})^2/\int J^2$, i.e.\ by the `angle' between K\"ahler 
form and flux. This angle can become small if the K\"ahler moduli are 
appropriately stabilised. Hence the  $\ln(r)$ term can be suppressed 
{\it beyond} the generic loop-factor. As we will explain in more detail 
below, this extra suppression allows us to avoid cosmic string constraints 
{\it without} going to extremely small couplings and relying on 
corrections near the point of tachyon condensation, as suggested in \cite{Haack:2008yb}.

We now return in more detail to the relation of our proposal to previously 
discussed scenarios of brane inflation. Considering just the form of the 
potential in (\ref{pot}), our setting appears to be rather close to the 
D3/D7 scenario \cite{Dasgupta:2002ew} (see also \cite{Hsu:2003cy,
Koyama:2003yc,Firouzjahi:2003zy,Berg:2004ek,Hsu:2004hi,Watari:2004xh,
Dasgupta:2004dw,Chen:2005ae,McAllister:2005mq,Burgess:2006cb,
Brandenberger:2008if,Haack:2008yb,Dasgupta:2008hw,Chen:2008au,
Burgess:2008ir,Gwyn:2010rj}). However, as anticipated before, the 
D7/D7 scenario has a much larger field range of the {\it canonically 
normalised} inflaton. Furthermore, we reiterate that for $\int {\cal F} \wedge {\cal F} =0$ appropriate K\"ahler moduli stabilisation 
allows us to flatten the potential beyond the generic $D$-term inflation 
case by the above argument. 

Next, we note that our setting is T-dual to scenarios where inflation arises
from branes at angles \cite{GarciaBellido:2001ky,Blumenhagen:2002ua,Jones:2002cv,
GomezReino:2002fs,Brandenberger:2003py,Epple:2003xt,Sato:2003kq,
Jones:2003ew,Shandera:2003gx,Matsuda:2004bk,Dutta:2007cr} (see 
\cite{Herdeiro:2001zb,Kyae:2001mk} for related earlier proposals). This 
is apparent if one thinks of a toroidal compactification and performs a 
T-duality along one of the two radii supporting the flux. While the inflaton 
is still the brane separation modulus, reheating now corresponds to
brane-recombination. Compared to scenarios with branes-at-angles, 
D7/D7 inflation has the advantage that it arises naturally in the 
(arguably) best-understood region of the landscape, i.e.\ in Type-II-B/F-theory 
with fluxes \cite{Giddings:2001yu,Kachru:2003aw,Denef:2004ze}. In particular, 
we hope that D7/D7 inflation can be investigated explicitly in rather generic 
geometries, and issues like the fine-tuning of the present-day cosmological 
constant through fluxes can be addressed simultaneously. 

Alternatively, one can perform a T-duality along one of the radii which
are not wrapped by the branes. This leads to Wilson line inflation
\cite{Avgoustidis:2006zp,Avgoustidis:2008zu}. In fact, the authors of 
\cite{Avgoustidis:2006zp} briefly mention the possibility of a T-dual 
version, where their Wilson line is replaced by the brane position. They 
emphasize the danger of a too steep, flux-induced potential. We will return 
to this critical issue in detail in Sect.~\ref{Modulistabilisation} and 
App.~\ref{app_moduli}, arguing that at leading order such a potential 
does not arise in appropriate geometries and for suitable fluxes. At 
subleading order, a flux-induced potential for the inflaton may actually be 
a crucial ingredient, which is necessary for a moderate tuning after K\"ahler 
moduli stabilisation. 

Our paper is organized as follows. In Sect.~\ref{sec_geom}, we start by 
describing the geometric setup and deriving geometric constraints which 
ensure that 7-brane inflation can work in the way outlined in the 
introduction. We then continue in Sect.~\ref{D7onT6} by calculating, both 
in 10d supergravity and by a string-theoretic one-loop analysis, the 
D7/D7-brane potential in the simplified case of a $T^6$ geometry. In App.~\ref{corr} we give a field-theoretic interpretation of the inter-brane potential in terms of loop-corrections to the FI-term and to the gauge kinetic function. The 
potential is then generalised to the Calabi-Yau situation in 
Sect.~\ref{Pot-gen}. A preliminary phenomenological analysis, including 
the CMB spectrum and cosmic string constraints, is the subject of 
Sect.~\ref{pheno}. 
We conclude, in Sect.~\ref{Modulistabilisation},
with an outlook on the specifics of competing $F$-term effects arising after K\"ahler moduli 
stabilisation.
A number of technical calculations are collected in the appendices.

\section{The geometric Calabi-Yau setup for 7-brane hybrid 
inflation}\label{sec_geom}

\subsection{The general mechanism}

In this section we describe the geometric configuration underlying our 
fluxbrane inflation scenario. We consider a general Type IIB orientifold 
compactification on a Calabi-Yau 3-fold $X_3$ modded out by the orientifold 
action $\Omega (-1)^{F_L} \sigma$. The  holomorphic involution $\sigma$ is 
chosen such that it gives rise to O3- and O7-planes compatible with the 
addition of D3- and D7-branes. The four-dimensional effective action of 
such Type IIB orientifolds has been studied in detail in 
\cite{Grimm:2004uq,Jockers:2004yj,Jockers:2005zy,Haack:2006cy}.

The key players of our inflationary scenario are spacetime-filling D7-branes 
wrapping holomorphic 4-cycles of $X_3$. The inflaton is related to the position 
modulus for a particular D7-brane as follows: In flat space, two parallel 
7-branes can be separated from each other in such a way that there is a 
non-zero distance $r$ in the perpendicular complex plane at every point of the 
branes. This means that there exists a modulus $y$ associated with the brane 
separation such that $|y| =r$. As we will discuss in detail in subsection \ref{sec-GeomConstr}, this simple picture 
receives interesting modifications for curved branes on general 
manifolds. Nonetheless one can maintain the concept of relative 7-brane motion 
 and of an associated modulus - the inflaton. Until Sect.~\ref{sec-GeomConstr} we will ignore all complications occurring on curved spaces as compared to flat backgrounds.

Concretely let us denote by $\Sigma \in H_4(X_3, \mathbb Z)$ a divisor class 
with a geometric deformation modulus; i.e.\ a 7-brane wrapped along a 
representative in the class $\Sigma$ can move in $X_3$. We 
assume for simplicity that this brane does not intersect the 
orientifold plane or its orientifold image in the class $\sigma^*\Sigma$. 
A pair of 7-branes ${\cal D}_a$ and ${\cal D}_b$ along two 
representatives $\Sigma_a$, $\Sigma_b$ of the divisor class $\Sigma$ can then be deformed with respect to one another.
\par

Our second ingredient is non-supersymmetric relative gauge flux along the two branes and the resulting attractive $D$-term potential. If the two branes are separated from each other, the 
four-dimensional gauge group is $U(1)_a \times U(1)_b$. On the D7-branes we 
switch on non-trivial $U(1)$ gauge bundles $L_a$ and $L_b$ with first 
Chern class 
\bea
c_1(L_a) = \frac{1}{2\pi} ( \ell_s^2  {F}_a ) + \iota^* B_+ 
\in H^2(\Sigma_a, {\mathbb Z}/2)\,,    
\eea
and analogously for $L_b$. Here we distinguish between the (dimensionful) 
expectation value of the curvature $F_i = dA_i$ and the pullback\footnote{In the following we will omit writing $\iota^*$ explicitly all the time. It will be clear from the context whenever we need to pull back a form to $\Sigma$.} with respect to the embedding $\iota: \Sigma \rightarrow X_3$ of a 
discrete $B$-field described by elements of $H^{1,1}_+(X_3)$ that are even 
under the orientifold involution $\sigma$. By contrast the quantity
\bea
 {\cal F}_a =   \frac{1}{2\pi} ( \ell_s^2  {F}_a )  +  B
\eea
refers to the full $B$-field including its non-integer piece along orientifold odd
elements of  $H^{1,1}_-(X_3)$. In our conventions the string length $\ell_s$ is related to the Regge slope $\alpha'$ as $\ell_s = 2 \pi \sqrt{\alpha'}$.

The dynamics of the relative brane motion during inflation involves only the relative gauge group $U(1)_{-}$ with abelian generator 
$ Q_- = \frac{1}{\sqrt{2}} (Q_a - Q_b)$.
In general there will be open strings stretched between ${\cal D}_a$ and  ${\cal D}_b$ charged under $U(1)_-$. In their ground state sector they give rise to chiral multiplets $\Phi^i_{ab}$ with charge $(-1_a, 1_b)$ and $\tilde \Phi^j_{ab}$ with charge $(1_a, - 1_b)$. In flat space, as the parallel branes are separated, these strings would necessarily acquire a supersymmetric mass term proportional to the brane separation with modifications on curved spaces to be discussed below.
Apart from appearing with a supersymmetric mass the bosonic components of $\Phi^i_{ab}$ and $\tilde \Phi^i_{ab}$ enter the four-dimensional  ${\cal N}=1$ supergravity $D$-term potential for $U(1)_{-}$ of the standard form\footnote{We use the same symbol $\Phi$ to denote the scalar component of a chiral superfield $\Phi$. It will always be clear from the context to which of the two we are referring.}
\bea
\label{D-term1}
V_D =  \frac{1}{2}\Re (f)^{-1}  \left(   -  \sqrt{2}\sum_i |\Phi^i_{ab}|^2 +  \sqrt{2}\sum_j    |\tilde \Phi^j_{ab} |^2 +   \xi_{ab}   \right)^2.
\eea
Here $f$ represents the gauge kinetic function  associated with the four-dimensional gauge group $U(1)_-$.
To first order its real part is given by 
\bea
\Re (f)
 =    
 \frac{1}{2 \pi}      \left(\frac{1}{2} \int_\Sigma  \hat J \wedge   \hat   J   -  e^{-\phi} \int_{\Sigma} \frac{1}{2} {\cal F}_{ab} \wedge   {\cal F}_{ab}       \right).
\eea
Here $\hat J \in H^2(X_3)$ is the K\"ahler form on $X_3$ as appearing in the ten-dimensional Einstein frame. It is related to the K\"ahler form $J$ in the ten-dimensional string frame via
\bea
\label{Kaehler-Einstein}
\hat J = e^{-\phi/2} J 
\eea
and is normalised such that $ {\cal V}(\Sigma) =\frac12 \int_{\Sigma}   J \wedge    J$ is dimensionless and measures the string frame volume of the divisor $\Sigma$ in units of the string length $\ell_s $. Furthermore, we have defined the relative flux as $ {\cal F}_{ab}  = \frac{1}{\sqrt{2}} (   {\cal F}_{a}  -   {\cal F}_{b})$.

The quantity $\xi_{ab}$
is known, by slight abuse of nomenclature, as the field-dependent Fayet-Iliopoulos term and 
serves as an order parameter for the amount of relative supersymmetry breaking. 
In the above conventions we have \cite{Haack:2006cy}
\bea
\label{FI1}
\xi_{ab} =  \frac{M^2_P}{4 \pi}   \frac{   \int_\Sigma    \hat J   \wedge   {\cal F}_{ab}  }  {   \hat {\cal V}(X_3)  }, \quad\quad \hat {\cal V}(X_3) = \frac{1}{6}  \int_{X_3}   \hat J \wedge \hat J \wedge \hat J, \quad\quad M_P^2 = \frac{4 \pi}{\ell_s^2},
\eea
where $M_P$ denotes the four-dimensional reduced Planck mass.

If the two branes are separated and $\xi_{ab} \neq 0$, an attractive potential between the branes arises. 
We will compute the precise form of this potential in Sect.~\ref{D7onT6} and Sect.~\ref{Pot-gen}.
Clearly the amount of supersymmetry breaking responsible for this potential depends dynamically on the K\"ahler moduli appearing in (\ref{FI1}). Thus stabilisation of the K\"ahler moduli in a non-supersymmetric manner is key to a successful realisation of inflation. For now, however, we postpone the question of moduli stabilisation and treat $\xi_{ab}$ as a parameter. 

The end of inflation is marked by the critical distance $r_{\rm crit.}$ at which one of the fields $\Phi^i_{ab}$ or $\tilde \Phi^j_{ab}$ becomes tachyonic.
To determine when this happens we must take into account, in addition to  the supersymmetric mass term for the string modes proportional to the brane distance, the non-supersymmetric mass from the $D$-term potential. To arrive at an expression for $r_{\rm crit.}$ we consider the case of a compactification on a factorisable torus $\prod_{I=1}^3 T^2_I$, reserving modifications on the curved backgrounds for the next subsection. Separating the two branes ${\cal D}_a$ and ${\cal D}_b$ by a distance $r$, measured in units of $\ell_s$,  yields a supersymmetric mass square $(2 \pi / \ell_s)^2 r^2 $ of the open string states $\Phi^i_{ab}$. To quantify the non-supersymmetric mass, suppose that the relative flux 
density is non-vanishing on one torus only and parameterized by\footnote{Whenever we work on the torus we will use a basis of coordinates $x^j$ which run from zero to $2\pi R_j $ where $R_j$ is the radius of the corresponding $S^1$. For more details on these conventions see Sect.~\ref{D7onT6}.}
\begin{equation}
 \ccF^{ab}_{45} =\frac{1}{\sqrt{2}}\left(\ccF^{a}_{45} - \ccF^{b}_{45}\right) = \frac{1}{\sqrt{2}}\left(\tan\theta_{a} - \tan\theta_{b}\right).
\end{equation}
Then, by T-duality, one can use the familiar result of the branes at angles picture for the mass of the lightest state \cite{Arfaei1997, Polchinski:1998rr}
\begin{equation}
 m^2 = \frac{(2\pi)^2}{\ell_s^2}r^2 - \frac{2\pi \theta_{ab}}{\ell_s^2},
\end{equation}
where we assumed (without loss of generality) that moduli stabilisation has resulted in $\theta_{ab} \equiv \theta_a - \theta_b > 0$. From this expression one can read off that the lightest state becomes tachyonic at the critical distance
\begin{equation}
 r_{\rm crit.}^2 =  \frac{\theta_{ab}}{2\pi}.
\end{equation}
To obtain the corresponding expression in terms of a canonically normalised field $\phi \equiv |\Phi|$ (the inflaton) in four dimensions we use the relation to the eight-dimensional modulus $y$ \cite{Burgess:2001fx}\footnote{Note that the authors of \cite{Burgess:2001fx} use different conventions for the rescaling of the metric in order to go from ten-dimensional string frame to four-dimensional Einstein frame.}
\begin{equation}\label{CanNorm}
 \frac{\phi}{M_P} = r \, \sqrt{\frac{g_s}{4}\frac{\mathcal{V}(\Sigma)}{\mathcal{V}(X_3)}}, \quad\quad r \equiv |y| .
\end{equation}
With the help of (\ref{FI1}) it follows that
\begin{equation}\label{bifurcationpoint}
 \phi_{\rm crit.}^2 \simeq \frac{\xi_{ab}}{\sqrt{2}}
\end{equation}
for small $\theta_{ab}$. This is precisely the result one would obtain by embedding hybrid inflation from $D$-terms in $\mathcal{N} = 2$ supersymmetry where there is a relation between the trilinear coupling $\lambda$ in the superpotential and the gauge coupling $g_{\rm YM}$ of the form $\lambda^2 = 2 g_{\rm YM}^2$ \cite{Kallosh:2001tm, Kallosh:2003ux}. The four-dimensional mass squared of the tachyon in this model is given by
\begin{equation}
 m^2_{4D} = 2\,\Re(f)^{-1}\left(\phi^2 -  \frac{\xi_{ab}}{\sqrt{2}}\right).
\end{equation}

\par

Note that the $D$-term $\xi_{ab}$ acquires a higher correction proportional to ${\rm log}(r)$ which will be computed in App.~\ref{corr}. In the above expression for $r^2_{\rm crit.} $ this correction is neglected. Tachyon condensation leads to the formation of a bound state between ${\cal D}_a$ and ${\cal D}_b$ and break the gauge group $U(1)_a \times U(1)_b$ to $U(1)_{+}$ with generator $\frac{1}{\sqrt{2}} \, (Q_a + Q_b)$. This is typical of hybrid $D$-term inflation, where the condensing tachyon  $\Phi^i_{ab}$ plays the role of the waterfall field.
\par

In a more general setup on curved backgrounds there is no unambiguous definition of a distance 
$r$ between the $D$-branes. Instead, we can use the relation (\ref{CanNorm}) 
as a definition of $r$ in terms of the four-dimensional inflaton $\phi$ as will be discussed next.

\subsection{Constraints on generic Calabi-Yau backgrounds}
\label{sec-GeomConstr}

\begin{figure}
 \begin{center} 
  \includegraphics[width=0.6\textwidth]{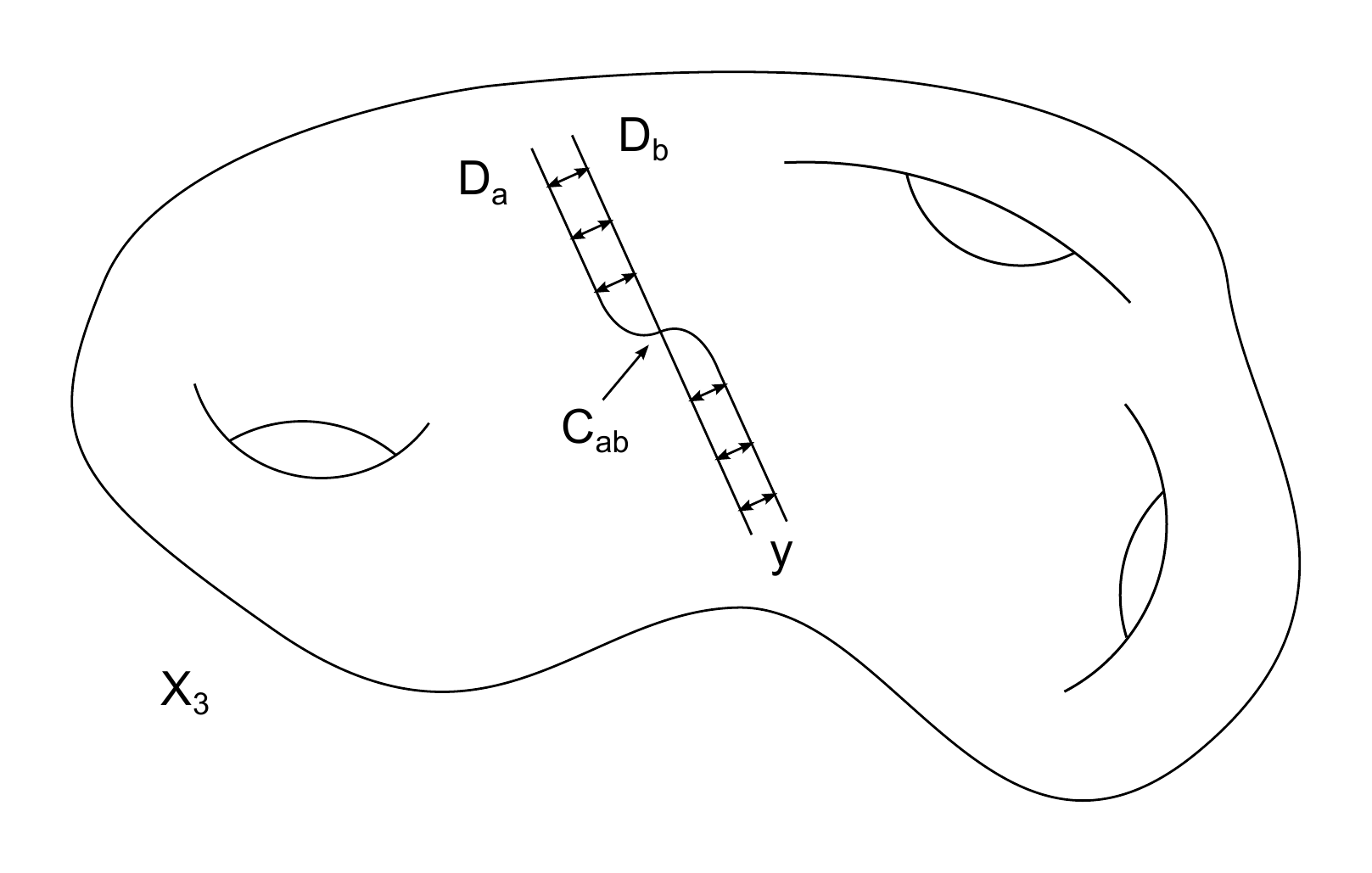} 
 \end{center} 
 \caption{\small Deforming homologous 7-branes on a Calabi-Yau.}\label{figa} 
 \end{figure}

In the previous section we used the notion of brane separation between two 7-branes along representatives $\Sigma_a$ and $\Sigma_b$ of a divisor class with deformation modulus. The simple picture discussed in the last subsection is modified on a general Calabi-Yau 3-fold $X_3$ because, in the presence of curvature, $\Sigma_a$ and $\Sigma_b$ 
cannot be separated from each other \emph{everywhere} along the 4-cycle.
Rather, they generically intersect along a  curve $C_{ab}$ given by the pullback of, say,  $\Sigma_b$ to $\Sigma_a$. This situation is represented in Fig.~\ref{figa}. 
 The self-intersection  curve $C_{ab}$ is in the homological class $[\Sigma]|_{\Sigma}$. The class $[\Sigma]$ is the first Chern class of the normal bundle $N_{\Sigma/X_3}$, which by adjunction equals the pullback of the canonical bundle $K_\Sigma|_{\Sigma}$.
Since generically $K_\Sigma|_{\Sigma}$ is non-zero, the self-intersection curve is non-trivial. 
In this case, a suitable deformation can still ensure that the two branes are at a large relative distance at points far away from the intersection curve, but there necessarily exists a region where the two branes come close to one another. For our applications this has two consequences: First we need to identify a good description of the \textit{effective} brane distance. Second we need to revisit the distance-dependent mass terms for strings stretched between the two branes and the appearance of tachyonic modes at the end of inflation.

To this aim we recall that deformations of a complex divisor are elements in $H^{(0,2)}(\Sigma) \simeq H^0(\Sigma, K_\Sigma)$,\footnote{In an orientifold, the deformations are in fact elements of the subspace $H^{(0,2)}_-(\Sigma)$ odd under the orientifold involution \cite{Jockers:2004yj}.} i.e.\ they are sections of the canonical bundle $K_{\Sigma}$. The self-intersection of $\Sigma$ corresponds to the fact that for non-trivial canonical bundle $K_{\Sigma}$, a section $\varphi \in H^0(\Sigma, K_\Sigma)$ necessarily vanishes along a curve on $\Sigma$. Since the value of $\varphi$ is a measure for the deformation of the two branes away from each other, it is therefore not possible to separate the branes everywhere.
Suppose for simplicity that $h^{(0,2)}(\Sigma)=1$ corresponding to a single section $\varphi \in H^{0}(\Sigma, K_{\Sigma})$. 
Due to the non-trivial profile of $\varphi$ along $\Sigma$ the separation of the representations $\Sigma_a$ and $\Sigma_b$ varies along the four-cycle.

Oriented by the considerations of the previous section the sought-after measure for the effective distance between  $\Sigma_a$ and $\Sigma_b$ is  the supersymmetric mass of strings stretched between  $\Sigma_a$ and $\Sigma_b$.
 This mass is set by the VEV of an ${\cal N}=1$ chiral superfield $\Phi(x)$ of the four-dimensional effective action which
arises by dimensional reduction of the eight-dimensional deformation modulus 
\bea
\zeta(x, z) = \Phi (x) \, \, \varphi(z), \quad\quad \varphi \in H^{0}(\Sigma,K_{\Sigma}).
\eea
To quantify the relation between $\langle | \Phi |\rangle$ and the supersymmetric mass of strings between  $\Sigma_a$ and $\Sigma_b$ we need to recall some general facts about the localisation of massless matter on D7-branes.

We must distinguish between so-called bulk states, i.e.\ modes propagating along the entire brane divisor, and  matter localised on the intersection curve of two D7-branes.
Consider first the situation of two coincident D7-branes along the divisor $\Sigma$ in the presence of supersymmetric gauge flux, $\xi_{ab} =0$.
In this situation the massless open string states are the bulk ground states  along the entire divisor $\Sigma$. These are counted by  cohomology groups with values in the line bundles $L_a$ and $L_b$ of the two branes.
More precisely, the number of massless ${\cal N}=1$ chiral  superfields $\Phi^i_{ab}$ with charge $(-1_a, 1_b)$ and, respectively, of  chiral superfields  $\tilde \Phi^j_{ab}$ with charge $(1_a, - 1_b)$ is counted by the dimension of extension groups ${\rm Ext}^1(\iota_* L_a, \iota_* L_b)$ and ${\rm Ext}^2(\iota_* L_a, \iota_* L_b)$ \cite{Katz:2002gh}. 
These extension groups in turn can be related to the following combinations of cohomology groups (see e.g.\ \cite{Blumenhagen:2008zz} for more information)
\bea
\label{Ext1}
\Phi^i_{ab}  \quad \leftrightarrow \quad {\rm Ext}^1 (\iota_* L_a, \iota_* L_b)&=& H^1(\Sigma, L_a\otimes L^\vee_b)+ 
  H^0(\Sigma, L_a\otimes L^\vee_b\otimes N_{\Sigma/X_3}), \nonumber \\  
\tilde \Phi^i_{ab} \quad  \leftrightarrow \quad  {\rm Ext}^2 (\iota_* L_a, \iota_* L_b)&=& H^2(\Sigma, L_a\otimes L^\vee_b)+ 
  H^1(\Sigma, L_a\otimes L^\vee_b\otimes N_{\Sigma/X_3})
\eea
with $N_{\Sigma/X_3}$ the normal bundle to $\Sigma$. 
On the other hand for two branes on two general 4-cycles $D_a$ and $D_b$ intersecting along a curve $C_{ab} = D_a \cap D_b$, massless matter arises on the intersection locus and the number of zero modes are counted by the cohomology groups
\bea
\label{Ext2}
 {\rm Ext}^1 (\iota_* L_a, \iota_* L_b) &=& H^0(C_{ab},L^\vee_a\otimes L_b  |_{C_{ab}} \otimes \sqrt{K_{C_{ab}}} ), \nonumber \\
  {\rm Ext}^2 (\iota_* L_a, \iota_* L_b) &=& H^1(C_{ab},L_a^\vee \otimes L_b  |_{C_{ab}} \otimes \sqrt{K_{C_{ab}}} )
\eea
with  $K_{C_{ab}}$ the canonical bundle of the matter curve $C_{ab}$.
In both cases the chiral index is given by the same expression, obtained via the Hirzebruch-Riemann-Roch theorem,
\bea
\label{index}
\chi =  - \# \Phi^i_{ab} +  \# \tilde \Phi^i_{ab}  =   - \int_{X_3} [D_a] \wedge [D_b] \wedge (c_1(L_a) - c_1(L_b)),
\eea
where $[D_a], [D_b]$ denote the 2-forms dual to the divisor classes of the two branes. I.e.\ in the first case $[D_a] = [D_b] = [\Sigma]$.

Suppose now we start with two coincident 7-branes ${\cal D}_a$ and  ${\cal D}_b$ with massless modes in the ${\cal D}_a-{\cal D}_b$ sector counted by (\ref{Ext1}).
Separating the two branes affects the massless modes.
In the four-dimensional effective action a supersymmetric mass term follows from a trilinear superpotential
\bea
W = c_{ij} \, \Phi \, \Phi^i_{ab} \, \tilde \Phi^j_{ab}
\eea
between the charged modes and the modulus $\Phi$ describing the relative brane distance.
The internal wavefunctions associated with these fields are, respectively, $\varphi \in H^{0}(\Sigma, K_{\Sigma})$, $\psi_{ab} \in  {\rm Ext}^1 (\iota_* L_a, \iota_* L_b)$ and $ \tilde \psi_{ab} \in  {\rm Ext}^2 (\iota_* L_a, \iota_* L_b)$, which we take to be normalised. The coupling matrix $c_{ij}$ is then the non-zero triple overlap of these wavefunctions. In particular, if the wavefunctions  $  \psi_{ab}$ and  $ \tilde \psi_{ab}$ have support only along the intersection curve, as is the case for the localised modes, this overlap vanishes because $C_{ab}$ is the zero-locus of the section $\varphi$.

 Thus, while in flat space all states $\Phi^i_{ab}, \tilde \Phi^j_{ab}$ continuously acquire a non-zero mass proportional to the brane distance,
  on curved space there is a chance that some of the charged fields remain massless by localising along the self-intersection curve $C_{ab}$. 
 Of course we know that if the spectrum is chiral, all chiral states must remain massless because the brane deformation cannot change the chiral index.
The question is, however, if also some of the vector-like pairs which may exist in addition to the chiral states remain massless by localising along the curve $C_{ab}$.
For \emph{generic} line bundles, the answer is no:
To decide how many of  $\Phi^i_{ab}, \tilde \Phi^j_{ab}$ get massive we must compute the dimensions of the cohomology groups $H^i(C_{ab},L_a\otimes L^\vee_b |_{C_{ab}} \otimes \sqrt{K_{C_{ab}}}), i=0,1$ and compare this to the spectrum before brane deformation, given by (\ref{Ext1}).  As we will detail momentarily, for generic line bundles the spectrum along the curve $C_{ab}$ is purely chiral, while the spectrum of bulk states generically contains vector-like pairs. The non-chiral matter therefore acquires non-zero mass upon brane deformation despite the self-intersection of the divisor $\Sigma$.

As anticipated, our measure for the effective brane distance is this supersymmetric mass.
Let us consider the simplest case of just a single pair of vector-like modes in (\ref{Ext1}) with superpotential $W = c_{11} \, \Phi\,   \Phi^1_{ab} \, \tilde \Phi^1_{ab}$.
Then in view of relation (\ref{CanNorm}) in the flat case, we can define the effective brane distance
 \bea
 \label{reff}
r_{\rm eff} =  {\cal C} \,   \langle |\Phi| \rangle, \quad\quad {\cal C} =  \frac{| c_{11} |}{M_P}  \sqrt{ \frac{4}{g_s }\frac{ \mathcal{V}(X_3)} {\mathcal{V}(\Sigma)} } .
 \eea
This quantity will allow us to generalise results for the inter-brane potential gained on flat backgrounds to curved compactification spaces.

Let us now revisit the appearance of tachyonic modes at the end of inflation. We aim at realising a $D$-term brane inflation scenario resulting in tachyon condensation at a critical brane separation $r_{\rm crit.}$. 
In those regions in K\"ahler moduli space where the line bundles induce a non-zero $D$-term $\xi_{ab}$, the resulting non-supersymmetric $D$-term mass leads to a non-degenerate spectrum in the ground state sector.  Let us assume that the K\"ahler moduli are stabilised in a regime where $\xi_{ab} > 0$ for definiteness. Then at non-zero brane distance, the bosonic fields $\Phi_{ab}^i$ become tachyonic, while the fields  $\tilde \Phi_{ab}^i$ have positive mass squared, see eq. (\ref{D-term1}). If in the supersymmetric case the fields $\Phi_{ab}^i$ become massive by the brane separation because they do not localise along $C_{ab}$, we are now in the situation that they will first become massless at a critical brane deformation $r_{\rm crit.}$ and then acquire positive mass square.

 In the inflationary context we must make sure that during inflation no tachyonic modes appear from the string ground states at the intersection curve $C_{ab}$, but rather that a massive string mode becomes tachyonic eventually.
  If we assume, as above, $\xi_{ab} > 0$ for definiteness, the condition for this is as follows:
 The modes localised along the curve $C_{ab}$ acquire only non-supersymmetric $D$-term masses from the potential (\ref{D-term1}), irrespective of the amount of brane deformation. Therefore we must ensure that no tachyons appear here, i.e.\ the number of modes $\Phi^i_{ab}$ along $C_{ab}$ must vanish,
 \bea
 \label{cohom2}
 H^0(C_{ab}, L_a \otimes L_b^\vee \otimes \sqrt{K_{C_{ab}}}) =0.
 \eea
 At the same time, we need a state $\Phi^i_{ab}$ whose internal wave function propagates along the whole divisor because these are the states which acquire both a supersymmetric mass term proportional to the brane separation and a non-supersymmetric $D$-term mass; they can therefore act as recombination moduli at the end of inflation. 
 We thus require that in addition
 \bea
 \label{cohom3}
 H^1(\Sigma, L_a\otimes L^\vee_b)+ 
  H^0(\Sigma, L_a\otimes L^\vee_b\otimes N_\Sigma) \neq 0.
 \eea
 These are two topological conditions which for generic line bundles are easily met.
 To see this, we recall that a generic line bundle  on a curve of genus $g$  has no sections if its degree is negative.
The line bundle appearing in (\ref{cohom2}) is $L_a \otimes L_b^\vee|_{C_{ab}} \otimes \sqrt{K_C}$. 
Let us introduce the notation $d= c_1(L_a \otimes L_b^\vee) |_{C_{ab}}$ for the degree of $c_1(L_a \otimes L_b^\vee) |_{C_{ab}}$.
From the Riemann-Roch theorem, $\int_{C_{ab}} c_1(K_{C_{ab}}) = 2g -2$, we can read off that the degree of $\sqrt{K_C}$ is $g-1$.
Thus, absence of states $\Phi^i_{ab}$ along $C_{ab}$ is guaranteed for generic line bundles as long as $d < 1 -g$.
On the other hand, (\ref{index}) teaches us that $d$ is also minus 1 times the chiral index $\chi$ of the states in the ${\cal D}_a - {\cal D}_b$ sector.
Consequently absence of tachyons along the curve $C_{ab}$ is guaranteed, for $\xi_{ab} > 0$, as long as $ \chi \geq g$ .
On the other hand, a generic line bundle configuration $L_a$, $L_b$ of positive chirality will give rise not to a purely chiral spectrum of states along the entire divisor, but include a set of vector-like pairs. In particular, $(\ref{cohom3})$ is generically satisfied.\footnote{In particular if we can tensor e.g.\ $L_a$ by any line bundle that restricts trivially to $C_{ab}$, the chiral index does not change. In general we can pick a suitable line bundle such that $(\ref{cohom3})$ holds.}

To conclude, despite the self-intersection of D7-branes it is possible to realise a brane inflation scenario where the end of inflation is marked by the recombination of modes that become tachyonic at a critical brane distance $r_{\rm crit.}$.

Finally, let us turn to the process of tachyon condensation itself. As will become apparent later, we are interested in ensuring that the vacuum energy due to $D$-term supersymmetry breaking be annihilated almost completely at the end of inflation. This implies that the resulting stable ground state after tachyon condensation should be supersymmetric.
In the process of tachyon condensation, the original gauge group $U(1)_a \times U(1)_b$ is higgsed by a non-zero VEV for, say, some $\Phi^i_{ab}$, to the diagonal subgroup $U(1)_+$.
The appearance of this gauge group can also be understood as follows: Before tachyon condensation,  the gauge field configuration on the divisor $\Sigma$ is described by a direct sum of the two line bundles $L_a \oplus L_b$ with structure group $U(1)_a \times U(1)_b$. 
As the notation (\ref{Ext1}) suggests, condensation of the tachyonic recombination moduli $\Phi^i_{ab}$ transforms this direct sum into a non-split \emph{extension} bundle $V$ described by the short exact sequence
\bea
0 \rightarrow L_a \rightarrow V \rightarrow L_b \rightarrow 0.
\eea
The structure group of this bundle $V$ is $U(2)$ so that $V$ breaks the would-be gauge group $U(2)$ for two coincident branes to the diagonal $ U(1)_+$.
Note that the Chern character of $V$ is given by ${\rm ch}(V) = {\rm ch}(L_a) + {\rm ch}(L_b)$. 
For the final configuration to be supersymmetric, two constraints must be met: An obvious necessary condition is that the $D$-term associated with the bundle $V$ must vanish. If $c_1(V)  = c_1(L_a) + c_1(L_b) \neq 0$ the  D-flatness condition depends on the K\"ahler moduli. Alternatively one can choose a configuration of gauge flux with $c_1(L_a) = - c_1(L_b)$ so that $c_1(V) =0$ and no $D$-term arises after recombination, independent of the K\"ahler moduli.
In addition, supersymmetry requires that  the non-abelian bundle $V$ must be $\Pi$-stable, a constraint that reduces, in the large volume limit, to the requirement of $\mu$-stability.\footnote{Recall that a rank $r$ vector bundle of slope $\mu$ is called $\mu$-stable if every coherent subsheaf of positive rank smaller than $r$ has slope $\mu' < \mu$. See e.g.\ \cite{Aspinwall:2004jr} and references therein. } This condition is considerably more involved and must be checked explicitly for a concrete choice of divisor class $\Sigma$ and line bundles $L_a$, $L_b$.

\section{Fluxed D7/D7-brane potential on {\boldmath $T^6$}}

\label{D7onT6}

In this section we present two alternative approaches to obtain the attractive potential of two magnetised D7-branes. Both computations will be carried out for the special case of a compactification on a factorisable six-torus ${T}_1^2 \times {T}_2^2 \times {T}_3^2$, postponing generalisations to genuine Calabi-Yau spaces to Sect.~\ref{Pot-gen}. In App.~\ref{corr} we provide yet another, field-theoretic interpretation of this potential.

The two D7-branes (called D$7_a$ and D$7_b$) wrap the first two tori $T^2_1$ and $T^2_2$ and are separated by a distance $r$, measured in units of $\ell_s$, on $T^2_3$. We will work in the limit in which the volume of $T^2_3$ goes to infinity. The torus $T^2_I$ has coordinates $x_{1+ 2I + j}$, $j = 1,2$, which run from zero to $2\pi R_j^I $ and the complex structures of the tori will be assumed to be purely imaginary (i.e.\ $u^I = i u^I_2 = i R^I_2 / R^I_1$).

\subsection{Potential from 10d supergravity perspective}
\label{probe-approx}

\begin{figure}
 \begin{center} 
 \includegraphics[width=\textwidth]{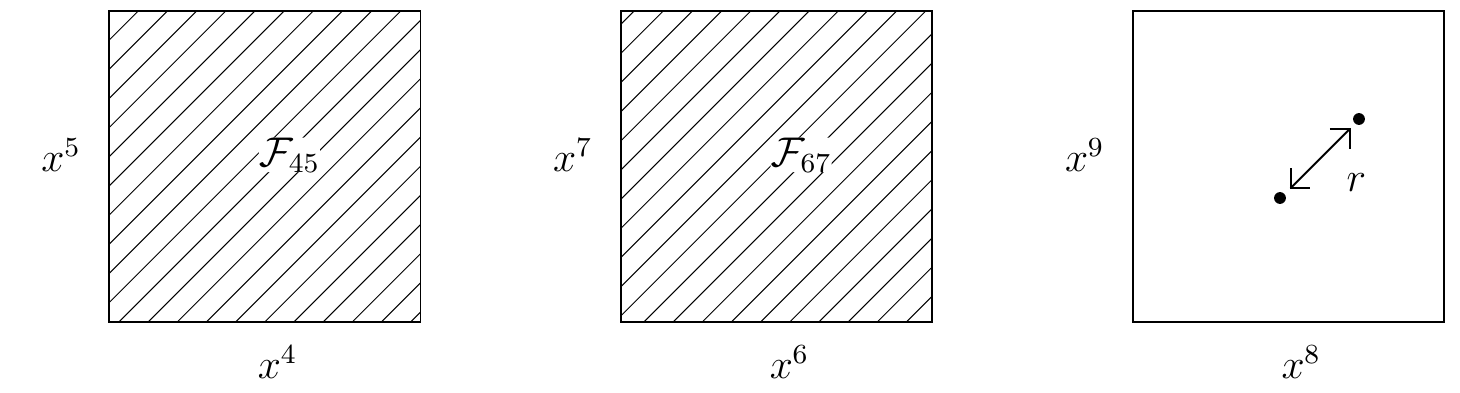} 
 \end{center} 
 \caption{\small Magnetised 7-branes on ${T}^2_1 \times {T}^2_2 \times {T}^2_3$.}\label{figb} 
 \end{figure}

We start by calculating the potential of a probe D7-brane with world-volume gauge flux $\mathcal{F}_a$ moving in the background of a D7-brane with world-volume gauge flux $\mathcal{F}_b$, roughly following \cite{Dasgupta:2002ew}. We allow for gauge flux on both tori wrapped by the D-branes, see Fig.~\ref{figb}.

Our starting point is the standard supergravity background solution of an extremal D5-brane \cite{Horowitz:1991cd,Youm:1997hw} rotated by angles in the relevant tori. This is T-dual to the fluxed D7-brane background solution we have in mind. Application of the Buscher rules \cite{Buscher:1987qj,Bergshoeff:1995as} leads to the NS-NS and R-R field profile for the D7-brane background solution \cite{Russo:1996if,Breckenridge:1996tt,Costa:1996zd}. We then calculate the potential for the probe brane moving in this background by evaluating the DBI and CS terms of the action.\par\medskip

Consider therefore the background of a stack of $N$ extremal D5-branes rotated by angles $\phi_b^1$ and $\phi_b^2$ on the first and second tori of the compactification manifold. The D5-branes source the $C_6$ field and curve spacetime such that in string frame
\begin{align}\label{D5metric}
  d\bar{s}^2&=Z_5^{-\frac{1}{2}}\left(-dx_0^2+\ldots + dx^2_3+ d\bar{x}^2_4+d\bar{x}^2_6\right)+Z_5^{\frac{1}{2}}\left(d\bar{x}^2_{5}+d\bar{x}^2_{7}+dx_8^2+dx_9^2\right)
\end{align}
with R-R-form potential $C_6$ and dilaton given by
\begin{align}\label{RRDilaton}
  C_{0123\bar{4}\bar{6}}=g_s^{-1}\left(Z_5^{-1}-1\right)\,,\qquad
  e^{2\phi}= g_s^2 Z_5^{-1}\,.
\end{align}
Here $x_1,\ldots,x_3,\bar{x}_4,\bar{x}_6$ denote the coordinates \emph{along the brane}, which are obtained by simple rotations on the respective planes from the \emph{torus} coordinates, and $\bar{x}_5,\bar{x}_7,x_8,x_9$ those perpendicular to it. Furthermore, the single-center harmonic function $Z_5$ on the transverse space is given by
\begin{equation}\label{D5harmonic}
 Z_5 = 1 + g_s N \left(\frac{1}{2\pi r_\perp}\right)^{2}
\end{equation}
with $(\ell_s r_{\perp})^2=\bar{x}_5^2+\bar{x}_7^2+x_8^2+x_9^2 $.
\par

We can then apply T-duality to transform this stack of D5-branes into a stack of D7-branes wrapping the first two tori with gauge flux density
\begin{equation}
\label{back-dense}
2\pi \alpha' F^b_{45} =  \tan\phi_b^1\,, \quad\quad   2\pi \alpha' F^b_{67} =  \tan\phi_b^2\,.
\end{equation}
The details of this computation, which involves the smearing of the 5-brane charge along the lines of  \cite{Peet:2000hn}, are presented in App.~\ref{app-derivation}.
The final D7-brane background metric reads
\begin{equation}
\label{D7-flux-background1}
ds^2=Z_7^{-\frac{1}{2}}ds^2(\mathbb{E}^{1,3})+Z_7^{-\frac{1}{2}}H_1 ds^2(\mathbb{E}^{2}_{45})+Z_7^{-\frac{1}{2}}H_2 ds^2(\mathbb{E}^{2}_{67})+Z_7^{\frac{1}{2}}ds^2(\mathbb{E}^{2}_{89})\,.
\end{equation}
Here we have introduced $H_{i}=\left(\cos^2\phi_b^i + \sin^2\phi_b^i Z_7^{-1}\right)^{-1}$ and line elements e.g.\ $ds^2(\mathbb{E}^{2}_{45})=dx_4^2+dx_5^2$. Furthermore, $Z_7$ is given by
\begin{equation}
\label{D7-flux-background2}
Z_7 = 1 - N \frac{g_s}{2\pi} \frac{1}{\cos\phi_b^1\cos\phi_b^2}\log\left(\frac{r}{R}\right)\, 
\end{equation}
with $R$ a typical radius of the compactification space, measured in units of $\ell_s$. As will become clear shortly it is chosen such that at $r = R$ the dilaton $\phi$ is normalised as $e^{\phi} = g_s$.
The fluxed D7-branes also source the remaining closed string fields.
For the Kalb-Ramond $B$-field the only non-vanishing terms one finds are
\begin{align}
B_{45} = -\tan\phi_b^1 + \tan\phi_b^1 Z_7^{-1} H_1\quad\text{and}\quad B_{67} =  -\tan\phi_b^2 + \tan\phi_b^2 Z_7^{-1} H_2\,.
\end{align}
One immediately sees that the $B$-field vanishes as $r$ approaches $R$ since there $Z_7\rightarrow 1$ and thus $H_{i}\rightarrow 1$. For D$p$-branes of lower dimensionality ($p<7$) this would correspond to a vanishing $B$-field at infinity. For $r\rightarrow 0$ the Kalb-Ramond field captures the flux given earlier in \eqref{back-dense}. Note that our result differs from  \cite{Dasgupta:2002ew} by an additional constant for the $B$-field, which is immaterial in their context but cannot be neglected in our calculation.
The dilaton is calculated to be 
\begin{equation}
 e^{2\phi}= g_s^2 Z_7^{-2} H_1 H_2\,,
\end{equation}
and the expressions for the RR-background fields are collected in  eq.~(\ref{C-field}) in App.~\ref{sugra}.

Equipped with the IIB supergravity background of a D7-brane (from now on we set $N=1$) with gauge flux  density (\ref{back-dense}),  we are prepared to compute the potential felt by a probe D7-brane with flux density $ 2 \pi \alpha' F^a_{45}=\tan{\phi_a^1}$ and $ 2 \pi \alpha' F^a_{67}=\tan{\phi_a^2}$ when it moves in this background. The potential follows by evaluating the  DBI and CS parts of the probe brane action in the background. Let us first work in the limit of essentially infinite internal dimensions, commenting on compactification effects  at the end of this section. Our conventions for the brane action in string frame are
\bea
 S_{\operatorname{DBI}} &=&-T_7 \int d^8 \sigma e^{-\phi} \sqrt{-\det\left(g_{\mu\nu}+\ccF_{\mu\nu}\right)}\,, \\
  S_{\operatorname{CS}}&=&\mu_7 \int \sum_i  C_i \wedge e^{ B + 2 \pi \alpha ' F}\,, \quad T_7 = \mu_7 =\frac{2\pi}{\ell_s^8}\,,
\eea
where the integrals are over the world-volume of the probe brane and the embedding via $\iota$ into ten-dimensional spacetime is left implicit. The probe brane shall be parallel to the background brane (static gauge). Adding the probe brane flux and the contribution from the $B$-field, which incorporates the background brane flux, we find for the gauge flux density on the probe brane
\begin{equation}\label{inducedFlux}
\begin{split}
 \ccF^{a}_{45}&=\tan\phi_a^1-\tan\phi_b^1+\tan\phi_b^1 Z_7^{-1} H_1\,,\\
 \ccF^{a}_{67}&=\tan\phi_a^2-\tan\phi_b^2+\tan\phi_b^2 Z_7^{-1} H_2\,.
\end{split}
\end{equation}
For the DBI part of the action and to lowest order in the flux densities one obtains
\begin{align}
  S_{\operatorname{DBI}}&=-T_7 \int d^8 \sigma e^{-\phi} \sqrt{-\det\left(g_{\mu\nu}+  \mathcal{F}_{\mu\nu}\right)}\nonumber\\
  &\begin{aligned}\approx&- \frac{2\pi }{\ell_s^4}\,\int d^4x\;\mathcal{V}_\parallel \,(\cos\phi_a^1\cos\phi_a^2)^{-1} \,g_s^{-1}\\
  & -\frac{1}{2}\frac{ 1}{\ell_s^4}\,\int d^4x\;\mathcal{V}_\parallel\log\left(\frac{r}{R}\right) \,\frac{\cos^2(\phi_a^1-\phi_b^1)+\cos^2(\phi_a^2-\phi_b^2)}{\cos(\phi^1_a)\cos(\phi^1_b)\cos(\phi^2_a)\cos(\phi^2_b)}\,,\end{aligned}
\end{align}
where $\mathcal{V}_\parallel$ is the internal volume of the probe D-brane, measured in units of $\ell_s$, after having stripped off the warping factors. Furthermore, we have omitted the kinetic terms of the brane scalars and higher curvature contributions. We moreover derive the CS parts of the action,
\begin{align}
  S_{\operatorname{CS}}&=\mu_7 \int \sum_i  C_i \wedge e^{ B + 2 \pi \alpha ' F}\nonumber\\
&=\frac{1}{\ell_s^4}\int d^4x\;\mathcal{V}_\parallel\log\left(\frac{r}{R}\right)\cdot\left[1+\tan\phi_b^2\tan\phi_a^2+\tan\phi_b^1\tan\phi_a^1+\tan\phi_b^1\tan\phi_b^2\tan\phi_a^1\tan\phi_a^2\right]\nonumber\\
&= \frac{1}{\ell_s^4}\int d^4x \;\mathcal{V}_\parallel\log\left(\frac{r}{R}\right) \cdot\frac{\cos(\phi_a^1-\phi_b^1)\cos(\phi_a^2-\phi_b^2)}{\cos(\phi^1_a)\cos(\phi^1_b)\cos(\phi^2_a)\cos(\phi^2_b)}\,,
\end{align}
where in the second line the charges of the dissolved D5-branes ($\propto\int \mathcal{F}$) and D3-branes ($\propto\int \mathcal{F}\wedge \mathcal{F}$) within the interacting flux D7-branes are manifest.

Expanding the whole action for small flux densities we deduce that the potential is given by
\begin{equation}
\label{pot-torus}
 V_{\operatorname{sugra}}(r)=\frac{2\pi }{\ell_s^4}\,\mathcal{V}_\parallel \,(\cos\phi_a^1\cos\phi_a^2)^{-1} \,g_s^{-1} + \frac{1}{2}\frac{\mathcal{V}_\parallel }{\ell_s^4} \frac{\left[\cos(\phi_a^1-\phi_b^1)-\cos(\phi_a^2-\phi_b^2)\right]^2}{\cos(\phi^1_a)\cos(\phi^1_b)\cos(\phi^2_a)\cos(\phi^2_b)} \log\left(\frac{r}{R}\right)\,. 
\end{equation}
We see that the supersymmetric configurations of $\phi_a^1-\phi_b^1=\pm\left(\phi_a^2-\phi_b^2\right)$ both imply a vanishing potential accounting for the BPS nature of these states. On the torus $T^4$ this corresponds to (anti-)self-dual flux \cite{Berkooz:1996km,Balasubramanian:1996uc} as shown in more detail in App.~\ref{SD_ASD_F}. 
Furthermore, we note that the $g_s$ dependence has vanished in the distance dependent part of the potential. This indicates that it arises as a one-loop effect from the open string sector. 
Most importantly, in \eqref{pot-torus} we explicitly see that the logarithmic term appears with a prefactor which roughly scales like $\sim\phi^4$ or $|\ccF|^4$ for small angles (cf.~\eqref{back-dense}). This can be parametrically small for large internal brane volume and order one integrated flux $\int \ccF \in \mathbb{Z}$. This very point, together with the parametrically different behaviour of the constant, implies the flatness of the inflationary potential motivated in the introduction.

The constant in \eqref{pot-torus}, which is the energy density of the probe brane, has to be supplemented by the energy density of the background brane
\bea
\frac{2\pi }{\ell_s^4} \,\mathcal{V}_\parallel \,(\cos\phi_b^1\cos\phi_b^2)^{-1}\,g_s^{-1}\,.
\eea
Moreover, the energy density of the final BPS state \cite{Uranga:2002ag} has to be subtracted to obtain the correct tachyon potential \cite{Sen:1999mg},
\bea\label{pot_const}
V_0 \approx \frac{1}{4} \,\frac{2\pi}{\ell_s^4} \,\mathcal{V}_\parallel \, \left[(\phi_a^1-\phi_b^1)-(\phi_a^2-\phi_b^2)\right]^2\,g_s^{-1} .
\eea
For details on the constant in the T-dual picture see \cite{GomezReino:2002fs}.

The above results are modified if one takes into account compactness of the internal manifold. Compactification effects have been discussed in the context of inflation from branes at angles e.g.\ in \cite{Shandera:2003gx}. It was found that these effects are subject to the same relative suppression with respect to the constant in the potential as the logarithmic term above and are subleading, in the regime of interest, towards the logarithm. This conclusion equally applies to our scenario so that we will ignore their effect in the sequel.

\subsection{Potential via one-loop string computation}\label{1loop}

In this subsection the inter-brane potential is calculated via a string computation. We explicitly evaluate the amplitude of a tree-level closed string exchange between the two magnetised D7-branes. The purpose of this subsection is twofold: Not only do we attempt to reproduce the form of the potential (\ref{pot-torus}) found in Sect.~\ref{probe-approx}, but we also try to show that (\ref{pot-torus}) remains valid for $r<1$, as long as $r$ is not too small. We will find that the lower bound on $r$, for which (\ref{pot-torus}) is a valid approximation, is parametrically given by the distance at which the lowest lying state in the open string spectrum becomes tachyonic. This is a crucial result because, as we will derive in Sect.~\ref{pheno}, D7/D7 inflation takes place precisely in the regime where $r<1$.
\par

We will work in the setup described at the beginning of this section and we set the Kalb-Ramond field $B$ to zero.
To break supersymmetry we turn on gauge bundles for the $U(1)_a$ and $U(1)_b$ gauge theories on $T^2_1$ which are parameterized by\footnote{For simplicity we concentrate here on the case where non-trivial gauge flux is living on $T^2_1$ only. The generalisation to a setup with non-trivial gauge flux on both $T^2_1$ and $T^2_2$ is straightforward and will be given at the end of this section.}
\bea
\ccF^a_{45} = \frac{p_a \ell_s^2}{(2\pi)^2 R^1_1 R^1_2} = \tan \phi_a \label{flux_quantisation_1} , \quad\quad
\ccF^b_{45} = \frac{p_b \ell_s^2}{(2\pi)^2 R^1_1 R^1_2} = \tan \phi_b \label{flux_quantisation_2} ,
\eea
where $p_{a,b}$ are the first Chern numbers of the gauge bundles, i.e.\ $\int c_1(L_{a,b}) = p_{a,b}$. This type of compactification was discussed in \cite{ Blumenhagen2000, Blumenhagen2000a,Blumenhagen:2000vk, Blumenhagen2001b}.
\par

We consider the limit in which $T^2_2$ and $T^2_3$ are large in units of the string length. This allows us to neglect the Kaluza-Klein tower (along $T^2_2$) and the winding modes (along $T^2_3$) of the open string. Furthermore, we will assume that any further D-branes and O-planes which we have to introduce in order to obtain a globally consistent model are far away from the two D7-branes that drive inflation. In this way, these additional objects will not alter the results obtained in this subsection significantly. The object of interest is thus the amplitude of a tree-level closed string exchange between D$7_a$ and D$7_b$ which, by world-sheet duality, is equal to the annulus amplitude of open strings that stretch between D$7_a$ and D$7_b$. The latter is given by (see e.g.\ \cite{Blumenhagen2000a})\footnote{The definitions and some useful relations of the modular functions are collected in App.~\ref{app_modfunc}.}
\begin{align}\label{loop-amplitude}
 \ccA_{ab} = \frac{- i \mathcal{V}_{||}}{2^5 \ell_s^4}  &(\ccF^a_{45} - \ccF^b_{45}) \int_0^{\infty} \frac{dt}{t^4}\exp\left(- 2\pi t r^2 \right)\times\nonumber\\
&\times \sum_{\alpha, \beta \in \{0,1/2\}}\eta_{\alpha\beta} e^{i\pi\delta_{ab}(1-2\beta)}
 \frac{\vartheta \myfrac{\alpha}{\beta}(0,it)^3}{\eta(it)^9}\frac{\vartheta\myfrac{\alpha+\delta_{ab}}{\beta}(0,it)}{\vartheta\myfrac{1/2+\delta_{ab}}{1/2}(0,it)} .
\end{align}
The amplitude is normalised to the volume of our four non-compact dimensions. Furthermore, we have defined $\delta_{ab} \equiv \phi_{ab}/\pi \equiv (\phi_a - \phi_b)/\pi$, and $\eta_{\alpha\beta} = (-1)^{2\alpha+2\beta+4\alpha\beta}$. The prefactor comes from an integration over the momenta in the external directions and in $T^2_2$. The flux-dependent factor arises due to non-commutativity of the zero-modes of the string in the presence of a magnetic field, which gives rise to a multiplicity of the `Landau levels' \cite{Abouelsaood:1986gd}. The $r$-dependent exponential accounts for the $r$-dependent mass of the zero-modes of the string stretched between the branes. The sum over the string oscillator modes gives rise to the modular functions which are summed over the different spin structures. The result is then integrated over conformally inequivalent annuli.
\par

In App.~\ref{app_1loop} it is shown that as long as $r^2 \gg |\phi_{ab}|/(2\pi)$ and $|\phi_{ab}| \ll 1$ the annulus amplitude $\mathcal{A}_{ab}$ in (\ref{loop-amplitude}) is well approximated by
\begin{equation}\label{approx-integral}
 \ccA_{ab} \approx \frac{ \mathcal{V}_{||}}{2^5 \ell_s^4}   (\ccF^a_{45} - \ccF^b_{45}) \int_0^{\infty} \frac{dt}{t^4}\exp\left(- 2\pi t r^2 \right) (t\phi_{ab})^3 .
\end{equation}
Evaluating the integral gives a contribution from the $(ab)$-sector to the inter-brane potential of the form
\begin{equation}\label{IIB-potential-1angle}
 V^{\rm 1-loop}_{ab}(r) \approx \frac{ \mathcal{V}_{||}}{2^4 \ell_s^4} \phi_{ab}^4 \log\left(\frac{r}{R}\right) .
\end{equation}
Here $R$ is a cutoff introduced to regulate the divergent integral in (\ref{approx-integral}). Equation (\ref{IIB-potential-1angle}) has to be complemented with identical expressions for the $(ba)$-sector, giving a factor of two.
\par

For the more general case of flux on both $T^2_1$ as well as $T^2_2$ we obtain, after a similar calculation,
\begin{equation}\label{IIB-potential-2angles}
V^{\rm 1-loop}_{ab}(r) \approx \frac{\mathcal{V}_{||}}{2^4 \ell_s^4}\left(\left(\phi^1_a - \phi^1_b\right)^2 - \left(\phi^2_a - \phi^2_b\right)^2\right)^2 \ln\left(\frac{r}{R}\right) .
\end{equation}
Together with the identical result in the $(ba)$-sector this precisely matches the distance dependent term in (\ref{pot-torus}) in limit of small $|\left(\phi_a^I - \phi_b^I\right)|$.
\par

In summary, we have shown that the potential derived via a probe brane approximation in supergravity is reproduced by the full string calculation in the limit of small flux. Furthermore, the string calculation allows us to extend the range of validity for the inter-brane potential beyond the naive lower bound $r \gtrsim 1$. In fact, the logarithmic form of the potential with the characteristic $\sim |\ccF|^4$ coefficient remains valid as long as $r$ is parametrically larger than the distance at which the lowest lying open string mode becomes tachyonic.

\section{The brane potential on generic Calabi-Yau manifolds}
\label{Pot-gen}

We now generalise the toroidal D7/D7 flux-brane potential to the potential for a 7-brane configuration on a genuine Calabi-Yau 
orientifold $X_3$. This 
potential will then be expanded for small flux density. The 
resulting expression reduces, in the setup described in Sect.~\ref{D7onT6}, 
precisely to eq. (\ref{pot-torus}). Furthermore, we will interpret our 
result as the Coleman-Weinberg potential of a 4d gauge theory and compare
it to the analogous expression for D3/D7-brane inflation.

In the notation of Sect.~\ref{sec_geom} we wrap two 7-branes along the two 
homologous divisors $\Sigma_a$ and $\Sigma_b$, both in homology class 
$[\Sigma]$. We think of the brane along $\Sigma_a$ as the fluxed probe brane 
as in the toroidal example of Sect.~\ref{probe-approx}. Before 
incorporating the effect of the unfluxed background brane along $\Sigma_b$ 
we first consider the effective action along $\Sigma_a$ following 
the analysis of \cite{Haack:2006cy}. Neglecting higher curvature 
contributions, the bosonic part of the DBI action for a D7-brane with 
gauge flux $\cal{F}$ reads 
\begin{align}
 S_{\Sigma_a} = - \frac{2\pi}{\ell_s^8} \int_{\mathcal{M}_4} d^4 x \, e^{-\phi} 
\sqrt{-\det g_{(4)}}\,\sqrt{\det \left(1+2\pi\alpha^\prime g^{-1}_{(4)} F_{(4)}
\right)}\,\,\Gamma. \label{dbi}
\end{align}
The interesting dynamics is encoded in the expression
\begin{align}
 \Gamma=\int_{\Sigma_a}  \sqrt{\det(g_{\Sigma_a}
+\mathcal{F})}\,\label{gamma_F}.
\end{align}
This expression determines both the $D$-term potential and the gauge coupling, 
\begin{equation}
 V_D = \frac{2\pi}{\ell_s^8} e^{-\phi} \Gamma_{\mathbb{F}}\left(e^{-2\phi} 
\mathcal{V}\right)^{-2}, \quad   
g^{-2}_{\operatorname{YM}}\,\,=\,\,\frac{2\pi}{\ell_s^8} (2\pi \alpha^\prime)^2 
e^{-\phi} \Gamma_{\mathbb{F}}\,\,\simeq \,\,\frac{1}{2\pi} \left(\frac{1}{2}
\int_{\Sigma_a} \hat{J} \wedge \hat{J}  \right)\,,
    \label{D_term_potential}
\end{equation}
where we neglected subleading flux contributions to the latter.

The standard BPS calibration conditions for D7-branes \cite{Marino:1999af} 
require $\mathcal{F}$ to be a $(1,1)$-form and furthermore
\begin{align}
 \frac{1}{2}\left( J + i \mathcal{F}\right)\wedge \left(J + 
i \mathcal{F}\right)=e^{i\theta} \sqrt{\frac{\det(g_{\Sigma_a}+
\mathcal{F})}{\det(g_{\Sigma_a})}} 
\operatorname{Vol}_{\Sigma_a}.\label{BPS_calibration}
\end{align}
If one takes the absolute value of the integral of \eqref{BPS_calibration}  and Taylor expands it for small 
flux density one finds, at quadratic order,\footnote{
In 
addition to this simple argument, Ref.~\cite{Haack:2006cy} also provides an 
alternative computation in their App.~B.
}
\begin{align}
 \Gamma \simeq \frac{1}{2}\int_{\Sigma_a} J \wedge 
J - \frac{1}{2}\int_{\Sigma_a} \mathcal{F} \wedge \mathcal{F}+
\frac{1}{2}\frac{\left(\int_{\Sigma_a} J \wedge \mathcal{F}
\right)^2}{\left(\frac{1}{2}\int_{\Sigma_a} J \wedge 
J \right)}+\cdots\,.\label{te}
\end{align}
These three terms correspond to the NS-NS tadpole, the flux-induced D3-brane 
charge and the leading-order $D$-term potential. As will become clear in 
a moment, they do not induce any force acting on the brane. 

The crucial term responsible for the motion of the brane arises at quartic order in $\mathcal{F}$ in the Taylor expansion in \eqref{te} and reads
\begin{align}
-\frac{1}{8} \frac{\left(\int_{\Sigma_a} J \wedge 
\mathcal{F}\right)^2}{\left(\frac{1}{2}\int_{\Sigma_a} J 
\wedge J \right)^3}\left[\left(\int_{\Sigma_a} J 
\wedge \mathcal{F}\right)^2-4 \left(\frac{1}{2}\int_{\Sigma_a} 
J \wedge J \right)\left( \frac{1}{2}\int_{\Sigma_a} 
\mathcal{F} \wedge \mathcal{F}\right)\right]\,.\label{CY_potential}
\end{align}

So far our brane is still a holomorphic divisor in an unperturbed 
Calabi-Yau orientifold. We need to generalise this expression to account 
for the background of the second, almost parallel, fluxless brane 
along $\Sigma_b$. In the toroidal model of Sect.~\ref{probe-approx} its
effect is to replace the flat metric by the warped expression
\begin{align}
 ds^2 = Z_7^{-\frac{1}{2}} dx_\parallel^2 + Z_7^{+\frac{1}{2}}dx^2_\perp\,, 
\label{torus_metric}
\end{align}
where $Z_7 = 1 - \frac{g_s}{2\pi}\log\frac{r}{R}$ and a corresponding 
dilaton profile arises.\footnote{
We 
note that this dilaton profile is exactly cancelled by the profile of 
$g_{(4)}$ in \eqref{dbi}, so it will play no role in what follows.
} 
Since in flat space the 7-branes are parallel, the effect of the warping 
on the probe 7-brane is accounted for by the substitution 
\bea
g_{\Sigma_a}\,\,\to\,\,Z_7^{-\frac{1}{2}} g_{\Sigma_a}\label{sub}
\eea
in the expression for $\Gamma$. The $r$ dependence is now encoded entirely in 
$Z_7$. We propose that the substitution \eqref{sub} captures the effect 
of the background brane even in the Calabi-Yau case. Clearly, this is 
only an approximation relying on the fact that, for sufficiently small $r$,
the branes are still locally (approximately) flat and parallel. 

To derive the flux-brane potential, we Taylor expand $\Gamma$ starting 
directly from \eqref{gamma_F}:
\begin{equation}
 \Gamma=\int_{\Sigma_a}  \sqrt{\det g_{\Sigma_a}}
\left\{1-\frac{1}{4}\mbox{tr}\left((g_{\Sigma_a}^{-1}{\cal F})^2\right)-
\frac{1}{8}\left[\mbox{tr}\left((g_{\Sigma_a}^{-1}{\cal F})^4\right)-\frac{1}{4}
\left(\mbox{tr}(g_{\Sigma_a}^{-1}{\cal F})^2\right)^2\right]+\cdots\right\}\,.
\label{gte}
\end{equation}
The first, ${\cal F}$-independent term scales as $Z_7^{-1}$ according to 
\eqref{sub}. However, the induced $r$ dependence is precisely cancelled 
by that of the CS-action discussed in Sect.~\ref{probe-approx}. In the 
$|\ccF|^2$ term, the $Z_7$ factors cancel so that no $r$ dependence is
induced. This is clearly the same cancellation already mentioned in the 
introduction. Together, these contributions correspond to those displayed in 
\eqref{te}. 

Finally, the crucial term, quartic in ${\cal F}$, comes with an overall factor $Z_7$
and hence induces a non-trivial $r$ dependence, as already discussed in 
Sect.~\ref{probe-approx} and App.~\ref{SD_ASD_F} in the toroidal case. The complication on general 
Calabi-Yau spaces comes from the variation of the inter-brane distance $r$
between the background brane along $\Sigma_b$ and the flux brane along 
$\Sigma_a$. More specifically, we have 
\begin{equation}
r\sim \phi \, \left\lVert \varphi(z) \right\rVert
\end{equation}
according to the discussion in Sect.~\ref{sec-GeomConstr}. Thus, the quartic terms in
${\cal F}$ of \eqref{gte} are accompanied by a factor 
\begin{equation}
Z_7=1-\frac{g_s}{2\pi}\log\frac{r}{R}=1-\frac{g_s}{2\pi}\left(
\ln\phi + \ln\left\lVert \varphi(z) \right\rVert +\cdots\right)
\end{equation}
{\it under} the $z$-integral integral. Fortunately, since we are only 
interested in the variation of the potential with $\phi$, the functional 
form of $\varphi(z)$ and the various constants are irrelevant. We immediately 
see that the $\phi$ dependence is obtained by simply multiplying the 
complete term quartic in ${\cal F}$ with an overall factor $-g_s/(2\pi)\ln(\phi/
\phi_0)$, where $\phi_0$ is an arbitrary normalisation.

With this understanding, we can return to the more elegant expression 
for the quartic ${\cal F}$ term given in \eqref{CY_potential}, incorporate the 
background effect as explained, and combine it with the leading term, quadratic in ${\cal F}$. The resulting $D$-term potential reads
\begin{align}
 V_D = \frac{1}{2} g^2_{\operatorname{YM}} \xi^2\left[1 +\frac{1}{4}  \left\{   
\frac{\left(\int_{\Sigma} J \wedge \mathcal{F}\right)^2}{\left(\frac{1}{2}
\int_{\Sigma} J \wedge J \right)^2}-4\,\frac{\left( \frac{1}{2}\int_{\Sigma} 
\mathcal{F} \wedge \mathcal{F}\right)}{ \left(\frac{1}{2}\int_{\Sigma} J 
\wedge J \right) }    \right\} \, \frac{g_s}{2\pi}\, {\rm log} \left( \frac{\phi}{\phi_0} \right)  
\right].
\end{align}
We propose this as a generalisation of the attractive potential found for 
branes at angles \cite{GarciaBellido:2001ky,GomezReino:2002fs} on a torus 
in Type IIA to a generic Calabi-Yau Type IIB orientifold.

As is well-known from the  general framework of hybrid inflation, the 
potential admits a field-theoretic interpretation  as a Coleman-Weinberg 
potential arising as a one-loop correction with the massive waterfall-fields 
running in the loop. Their masses are split after SUSY-breaking due to the 
non-vanishing FI-term $\xi$. 

To account for the one-loop nature \cite{Coleman:1973jx} of the 
Coleman-Weinberg term we make a factor of $g_{\operatorname{YM}}^2$ explicit 
via \eqref{D_term_potential} and find
\begin{align}
\label{D7-final}
 V_D = \frac{1}{2} g^2_{\operatorname{YM}} \xi^2\left[1+  
\frac{g^2_{\operatorname{YM}}}{16 \pi^2} \left\{\frac{\left(\int_{\Sigma} J 
\wedge \mathcal{F}\right)^2}{\left(\frac{1}{2}\int_{\Sigma} J \wedge J 
\right)}-4\cdot\left(\frac{1}{2}\int_{\Sigma}\mathcal{F}\wedge\mathcal{F}
\right)\right\}   {\rm log} \left( \frac{\phi}{\phi_0} \right)    \right].
\end{align}
Now the first term in the big round brackets proportional to $(\int_{\Sigma} J \wedge \mathcal{F})^2/\frac{1}{2}\int_{\Sigma} J \wedge J$ can be made parametrically small.
At the same time the gauge flux can in principle be chosen in such a way that the induced D3 charge $\int_{\Sigma}\mathcal{F}\wedge\mathcal{F}$ of our D7-brane vanishes. In such situations one arrives at a highly suppressed logarithmic term which specifically arises in our D7-brane context.

At this stage it is instructive to compare the inflationary brane potential with the setup in D3/D7 inflation. 
From  \cite{Haack:2008yb} we recall that the D3/D7 potential takes the generic form
 \begin{align}
  V = \frac{1}{2} g^2_{\operatorname{YM}} \xi^2\left(1+ \frac{g^2_{\operatorname{YM}}}{16 \pi^2} \log \frac{\phi}{\phi_0}\right).
 \end{align}
In particular, there is no analogue of  the term  proportional to $(\int_{\Sigma}
J \wedge \mathcal{F})^2/\frac{1}{2}\int_{\Sigma} J \wedge J$, which arises from the non-alignment of relative D5-brane charge.
Rather, the expression only involves the relative D3-brane charge of the fluxed D7 and the mobile D3-brane.
To match this with the D7/D7-potential (\ref{D7-final}) we note the general result (see e.g.\ \cite{Blumenhagen:2008zz} for details) that for gauge flux that can be made supersymmetric inside the K\"ahler cone the expression $-\int_{\Sigma}\mathcal{F}\wedge\mathcal{F}$
is positive and thus measures D3 (as opposed to anti-D3) charge.

\section{Phenomenological analysis}\label{pheno}

In this section we collect the basic phenomenological properties of the inflationary $D$-term potential between two D7-branes. As one of our main results we will show that 
the D7/D7 inflationary scenario provides a mechanism to overcome the clash with observational bounds due to cosmic string production at the end of inflation. In fact these bounds have turned out to be a notorious problem in $D$-term inflation models with an underlying $\mathcal{N} = 2$ structure \cite{Kallosh:2001tm}.\footnote{Recall the discussion below eq.~(\ref{bifurcationpoint}).}
\par

Our potential is of the general type
\bea
\label{pot1}
V(\phi) = V_0 \,  \left(1 + \alpha \, {\rm log} \, \frac{\phi}{\phi_0} \right)
\eea
for the canonically normalised inflaton field $\phi$. Here we used the parametrisation
\bea
V_0 = \frac{1}{2} g_{\rm YM}^2 \xi^2, \quad \quad \alpha = \frac{g_{\rm YM}^2}{16 \pi^2} \left(- 2 \int_\Sigma \mathcal{F}^2  + \frac{g_{\rm YM}^2}{2\pi} \left(\int_\Sigma  \hat{ J} \wedge {\mathcal{F}}\right)^2  \right).
\eea
The choice of $\phi_0$ corresponds to some choice of normalisation for the potential. Its value is irrelevant at our level of precision. For convenience we will choose $\phi_0$ such that it corresponds to the bifurcation point $\phi_0 \equiv \phi_{\rm crit.}$ of our potential, defined in (\ref{bifurcationpoint}), which is where the tachyon appears and inflation ends. Close to this point the simple functional form (\ref{pot1}) is no longer valid.
\par

Let us first  analyse the field range required to obtain $N=60$ e-foldings in the course of inflation. To this end we recall that during slow-roll inflation the Hubble parameter $H \equiv \dot a(t)/a(t)$ and the potential $V(\phi)$ are related via Friedmann's equation
\bea
 3 H^2 = V , 
\eea
while the equation of motion for the inflaton takes the form
\bea
  3 H \dot \phi = - V' .
\eea
In these expressions we have set the reduced Planck mass $M_P \equiv 1$ for convenience. The number of e-foldings  follows from the inflationary potential as
\bea
\label{efolds}
N = \int_{t_N}^{t_0}   dt \, H =  \int_{\phi_0}^{\phi_N} d \phi \, \frac{V}{V'},
\eea
where $t_N$ denotes the time associated with the onset of  the last $N$ e-foldings  and $t_0$ marks the end of inflation; the  corresponding values of the inflaton are $\phi_N \equiv \phi(t_N)$ and $\phi_0 \equiv \phi(t_0)$. In our model inflation starts out far from the bifurcation point of the potential (i.e.\ $\phi_N \gg \phi_0$). A simple parametrical analysis shows that in a regime with good validity of the supergravity approximation (i.e.\ the typical length scales of the compactification manifold are large in units of the string length), the constant of the potential dominates over the distance-dependent term throughout inflation, i.e.\ $ \alpha \log(\phi/\phi_0) \ll 1$. This allows us to evaluate (\ref{efolds}) as
\bea
\frac{V}{V'} = \frac{\phi}{\alpha} \Longrightarrow N = \frac{1}{2 \alpha} \left(\phi_N^2 - \phi_0^2 \right).
\eea
Using $\phi_N \gg \phi_0$ we thus find that the field value of the inflaton at the beginning of the last 60 e-foldings is given by
\bea
\label{phiN}
\phi_N  \simeq \sqrt{2 \alpha N}.
\eea
\par

The slow roll parameters are readily evaluated, in the approximation (\ref{phiN}),  as
\bea
&& \epsilon = \frac12  \left. \left(\frac{V'}{V}  \right)^2 \right|_{t=t_N} = \frac{1}{2} \frac{\alpha^2}{\phi_N^2} = \frac{\alpha}{4N}, \\
&& \eta = \left. \frac{V''}{V}\right|_{t=t_N} = - \frac{\alpha}{\phi_N^2} = - \frac{1}{2N} .
\eea
Since $\alpha \ll 1$ it follows that  $\epsilon \ll |\eta|$  and thus for $N=60$ the slow-roll condition $\epsilon \ll1$, $|\eta| \ll1$ is easily satisfied. 
From the above we extract a prediction for the spectral index $n_s$ via
\bea
n_s = 1 - 6 \epsilon + 2 \eta \simeq 1 + 2 \eta  = 1 - \frac{1}{N} = 0.983.
\eea
This value lies marginally outside the $1\sigma$ value $n_s = 0.968 \pm 0.012$ according to WMAP7~\cite{Komatsu:2010fb}.
\par

The inflationary potential is further constrained by measurements of the amplitude of adiabatic curvature perturbations. They set a value for the ratio $V^{3/2}/V'$ at time $t_N$ as~\cite{Komatsu:2010fb} 
\bea\label{COBEnorm}
\tilde \zeta \equiv \left. \frac{V^{3/2}}{V'}\right|_{t=t_N} = 5.4 \times 10^{-4} .
\eea 
Using the smallness of the distance-dependent term relative to the constant of the potential (\ref{pot1}), as discussed above, we can evaluate this constraint in the approximation (\ref{phiN}) and for $N = 60$ as
\bea
\label{constraint1}
 \frac{V_0}{\alpha} = \frac{\tilde \zeta^2}{2 N} = 2.4 \times 10^{-9}.
\eea
To appreciate the implications on the parameters of our potential it is more convenient to analyse the inverse combination
\bea
\label{constraint2}
\frac{\alpha}{V_0} = \frac{1}{(2 \pi)^2 \xi^2 } \left( \int_\Sigma -\mathcal{F}^2  \right) +  2   \frac{  \hat{\mathcal{V}}^2(X_3) }{ \hat{\mathcal{V}}(\Sigma) }  = 4.2 \times 10^{8}.
\eea
Crucially, the first summand involves the FI-term $\xi$. For positive $ \int_\Sigma -\mathcal{F}^2 $ this therefore sets a lower bound on $\xi$, which turns out to lie above the observational bound from cosmic strings. In particular, this is the situation encountered in D3/D7 inflation, where $ \int_\Sigma -\mathcal{F}^2$ is replaced by a positive order one number.
\par

Let us pause for a moment to review the origin of the cosmic string bound. Generically, cosmic strings will be produced at the end of brane inflation when the tachyon appears and the waterfall sets in \cite{Majumdar:2002hy, Jones:2002cv, Sarangi:2002yt, Jones:2003da, Dvali:2003zh, Dvali:2003zj, Copeland:2003bj}. If this happens, an impact on observable quantities such as the CMB power spectrum is expected.\footnote{For a recent discussion of these effects see e.g.\ \cite{Copeland:2009ga}.} However, the spectrum produced by cosmic strings shows the wrong behaviour to serve as the main source for the temperature anisotropies observed in the CMB. (It has one broad peak and falls off much slower at high multipole moment $l$ than the spectrum obtained from inflation. The latter is due to the fact that the string continues contributing to small scale anisotropies after recombination. In this way, these anisotropies do not suffer from Silk damping.) On the other hand, cosmic strings are not ruled out entirely by observations. They may contribute a small fraction to the overall power of CMB fluctuations. This possibility has been analysed numerically (see e.g.\ the recent \cite{Battye:2010xz}). The most important result of these analyses for us is that they constrain the cosmic string tension $\mu$ which is commonly quoted in terms of the dimensionless quantity $G \mu$ where $G$ is Newton's constant. The bounds that can be found in the literature tend to vary roughly from $2\times 10^{-7}$ to $7\times 10^{-7}$ depending on the kinds of methods that are used to simulate the evolution of the string network as well as the input dataset. These bounds correspond to a constraint for the contribution from cosmic strings to the total power of the CMB radiation which is roughly $\lesssim 10 \%$ of the total power at multipole moment $l = 10$. Simulations show \cite{Battye:2010xz} that such a small contribution does not change the value of the spectral index $n_s$ significantly at the $2\sigma$-level. We will work with the latest result from WMAP7 \cite{Komatsu:2010fb} ($n_s = 0.968 \pm 0.012$ at $1\sigma$) and with a value of
\begin{equation}\label{xi-constraint}
 G\mu \lesssim 6.4 \times 10^{-7}
\end{equation}
for the string tension. This was found in \cite{Battye:2010xz} using the Abelian-Higgs model to simulate the evolution of the string network.\footnote{There are alternative approaches to look for signatures of cosmic strings, the most recent of which include \cite{Brandenberger:2010hn, Christiansen:2010zi, Dunkley:2010ge}. However, they seem to find upper bounds on $G\mu$ which are comparable to the one cited above.
\par
After the first version of this paper had been submitted, the improved cosmic string analysis \cite{Urrestilla:2011gr} appeared, including in particular WMAP7 data. The authors report values for $G \mu$ which are lowered by a factor of $2-3$ (depending on the kind of small-scale data they take into account in addition to WMAP7) compared to the bounds used in the present paper. This does not change our conclusions qualitatively. If anything, it makes our specific mechanism for the suppression of cosmic strings more important.
} This value implies a contribution of $\lesssim 9.3 \%$ from cosmic strings to the total power in the CMB at multipole moment $l = 10$. The cosmic string tension $\mu$ is related to the FI-term $\xi$ as \cite{Dvali:2003zh} $\mu = 2\pi\xi$. Therefore, (\ref{xi-constraint}) puts an upper bound on the size of the FI-term.
\par

If one takes the simulations seriously at the $1\sigma$ confidence level one observes a trend towards higher values of $n_s$ if cosmic strings are included in the simulation \cite{Battye:2010xz}. This fact helps e.g.\ in the discussion of D3/D7 inflation models \cite{Haack:2008yb}.
\par

Note that the inclusion of cosmic string anisotropies in the CMB power spectrum will obviously lower the contribution from inflationary curvature perturbations (i.e.\ they will lower the value (\ref{COBEnorm})). However, this is a minor effect and completely irrelevant at our level of precision. Therefore, we will entirely neglect this subtlety.
\par

Returning to the phenomenological discussion of brane inflation we note that for D3/D7 inflation the value of $\xi$ required by the measured value for the amplitude of curvature perturbations (and determined via the D3/D7 analog of equation (\ref{constraint2})) lies above the cosmic string bound (\ref{xi-constraint}). By contrast, our D7/D7 inflation model is in a fundamentally different position. Namely, by a suitable choice of gauge flux it is possible to achieve $\int_{\Sigma} \mathcal{F}^2 =0$ so that the FI-term completely drops out from (\ref{constraint2}). In this situation, what is constrained by (\ref{constraint2}) is the ratio of the volume of $\Sigma$ and of the Calabi-Yau $X_3$,
\bea
\label{constraint3}
 \frac{1}{2} \int_{\Sigma} \hat J \wedge \hat J    \simeq 4.8 \times  10^{-9} \left( \frac{1}{6} \int_{X_3} \hat J \wedge \hat J  \wedge \hat J \right)^2.
\eea
From this constraint we can extract a prediction for the typical volume of the compactification manifold: If for simplicity we assume an isotropic internal space with typical length $\hat{R}$  measured in units of the Einstein-frame string length $g_s^{1/4} \ell_s$, this means $\hat{R} \simeq 10$, which leads to a regime with good validity of the supergravity approximation. It remains to be checked that this prediction leads to an overall consistent picture of D7/D7 inflation. The remainder of this section is devoted to this issue.
\par

Due to our choice of a flux vector $\ccF \in H^2(\Sigma)$ which satisfies $\int_{\Sigma} \ccF^2 = 0$, the FI-term is in principle unconstrained by (\ref{constraint2}) and thus the measured value for the amplitude of curvature perturbations is not in conflict with bounds on $\xi$ from cosmic strings: Let $\xi_{\rm crit.}$ be the `critical' value of the FI-term for which the cosmic string bound is saturated, i.e.\ $\xi_{\rm crit.} / 4  \simeq 6.4 \times 10^{-7}$. With the help of (\ref{constraint3}) one can re-express the cosmic string bound $\xi \lesssim \xi_{\rm crit.}$ as
\bea
\frac{\left(\int_\Sigma \hat J  \wedge \mathcal{F}\right)^2}{\frac{1}{2} \int_{\Sigma} \hat J \wedge \hat J  } \lesssim 16 \pi^2 \xi^2_{\rm crit.} \times 4.2 \times 10^{8} \simeq 0.4 \, .
\eea
This makes it obvious that for generic intersection forms on the divisor $\Sigma$ a suitable choice of flux vector $\mathcal{F}$ can well accommodate the cosmic string bound in a manner consistent with the prediction $\hat{R} \simeq 10$ deduced from the normalisation of the amplitude of curvature perturbations (\ref{constraint2}).
\par

We proceed by briefly discussing the implications of the above analysis for the field range during inflation. Recall that in all of the above we assumed that inflation starts far away from the bifurcation point, which is the point where inflation ends. This means that we have to require $\phi_0 \ll \phi_N$ and thus, in view of (\ref{phiN}),
\begin{equation}\label{constraint4}
 \phi_0^2 \ll 2\alpha N .
\end{equation}
According to the discussion in Sect.~\ref{sec_geom} the bifurcation point is just $\phi_0^2 \simeq \xi/\sqrt{2}$. Assuming that $\int_{\Sigma} \mathcal{F}^2  =0$ and that the cosmic string bound (\ref{xi-constraint}) is saturated (i.e.\ $\xi \simeq \xi_{\rm crit.} = 2.6 \times 10^{-6}$) the requirement (\ref{constraint4}) can be rewritten as
\begin{equation}
 3.6\times 10^2 \ll \frac{\hat{ \mathcal{V}}^2(X_3)}{\hat{ \mathcal{V}}^2(\Sigma)} .
\end{equation}
This condition is in agreement with the prediction $\hat{R} \simeq 10$ for a typical length scale of our (isotropic) compactification manifold. We now have a consistent picture of D7/D7 inflation, described in its low-energy limit by $D$-term hybrid inflation, in the regime where $\phi_N \gg \phi_0$.
\par

As a final step we deduce from the above analysis the brane separation $r_N$ of the two D7-branes at the beginning of the last 60 e-foldings of inflation. The field value of the inflaton at this time is given by (\ref{phiN}). Considering for simplicity the case of a toroidal compactification, we may use (\ref{CanNorm}) to calculate $r_N$
\begin{equation}
 r_N^2 = 16 \pi N g_s^{-\frac{1}{2}} \xi^2 \frac{\hat{\mathcal{V}}^3(X_3)}{\hat{\mathcal{V}}^3(\Sigma)} \, ,
\end{equation}
where $r_N$ is measured in units of $\ell_s$ and $\xi$ is measured in units of $M_P$. Assuming again that the cosmic string bound is saturated (i.e.\ $\xi \simeq \xi_{\rm crit.} $) it is obvious that a roughly isotropic compactification manifold with typical length $\hat{R} \simeq 10$ measured in units of $g_s^{1/4} \ell_s$ leads to inflation in the regime where $r_N < 1$, more precisely it leads to $r_N^2 \approx 10^{-2}/\sqrt{g_s}$. This crucial conclusion makes it necessary to perform the full string computation in Sect.~\ref{1loop} in order to derive the inter-brane potential because, generically, the supergravity approximation can be trusted only at distances larger than the string length.

\section{Conclusions and outlook on moduli stabilisation and {\boldmath $F$}-terms}
\label{Modulistabilisation}

In this paper we have considered an inflationary scenario with 7-branes wrapping holomorphic divisors in Type IIB Calabi-Yau orientifolds. The role of the inflaton is played by the open string modulus describing the relative distance between two 7-branes along homologous 4-cycles. Non-supersymmetric gauge flux induces an attractive force between the two branes which results in a logarithmic potential for the inflaton. Field-theoretically this potential can be viewed as a brane-distance-dependent correction to the $D$-term of Coleman-Weinberg type. Our scenario  fits into the class of hybrid inflationary models and is dual to inflation from branes at angles. 

We have employed a two-fold strategy to compute the gauge-flux-induced inflaton potential. On the one hand we have performed a supergravity computation evaluating the action of a non-supersymmetric fluxed probe brane in the background of another brane. While this computation is performed on a toroidal background in the supergravity regime, we have been able to make a proposal, given in eq. (\ref{D7-final}), to generalise it to genuine Calabi-Yau spaces. Inflation ends at a brane distance where open strings stretched between the two branes become tachyonic. Since this distance lies considerably below the string scale, we have set out, again on a toroidal background, to perform an alternative, stringy computation of the potential by evaluating the one-loop potential in the open string channel. In fact we have established that the form of the potential at substringy distances above a certain flux-dependent lower bound agrees with the supergravity computation. We interpret this as evidence that also our proposal for the generalisation of the $D$-term  potential to non-toroidal Calabi-Yau spaces remains valid in the cosmologically relevant regime.

From a phenomenological perspective, one of the main virtues of our D7/D7-inflationary scenario is that already the logarithmic $D$-term potential by itself can avoid the tight cosmic string bounds which are notoriously problematic in the related D3/D7 scenario. Furthermore the inflaton range is easily compatible with the cosmologically required number of e-foldings.  We view this as encouragement to further pursue the implications of our 7-brane inflation scenario and to embed it in particular in the context of genuine F-theory compactifications.

Our analysis in this paper has focused on the form of the attractive $D$-term potential between separated 7-branes with non-supersymmetric flux and has shown that this $D$-term potential as it stands is of a type favourable for inflation. Of course the ultimate success or failure of an inflationary model hinges crucially upon the suppression or appearance of competing inflaton dependent terms in the full scalar potential; these may well spoil the originally envisaged inflationary mechanism. The appearance of such terms is intimately linked with moduli stabilisation.
A quantitative treatment of moduli stabilisation and of corrections to the $D$-term potential is beyond the scope of this article and relegated to future work \cite{wip}; here we merely summarise the form of competing $F$-terms and point out some of the most important challenges in the context of moduli stabilisation.

Our analysis has treated the $D$-term as a given order parameter for supersymmetry breaking. As stressed, however, the flux induced $D$-term in Type IIB orientifolds is well-known to depend dynamically on the K\"ahler moduli. Their stabilisation in a supersymmetry breaking regime is of pivotal importance for a successful inflationary scenario.
Note that this requirement is equally relevant for all variants of $D$-term inflation including the scenario of D3/D7 inflation or the T-dual inflation with branes at angles. In particular it is crucial not only to stabilise the overall Calabi-Yau volume, which has been in the focus of the literature so far, but the particular combination of K\"ahler moduli entering the $D$-term $\xi_{ab}$; this would be the prime candidate for a runaway direction that could spoil inflation. 
Our strategy in \cite{wip} will be to approach this in the context of large volume models \cite{Balasubramanian:2005zx,Cremades:2007ig}, where the volume modulus of the large cycle can be stabilised by $\alpha'$-corrections in the K\"ahler potential as opposed to non-perturbative superpotential terms. The latter are more difficult to use in this case because of the gauging of the K\"ahler modulus  and the associated necessity to take into account charged modes in the superpotential  \cite{Achucarro:2006zf,Blumenhagen:2006xt,Ibanez:2006da,Florea:2006si,Haack:2006cy,Cremades:2007ig,Blumenhagen:2007sm}.

A related challenge concerns the inevitable appearance of $F$-term contributions to the 7-brane modulus potential that may compete with the attractive and favourable $D$-term potential.
In fact, there are three qualitatively different sources for such a contribution: brane modulus dependent corrections to the K\"ahler potential, a direct appearance of the brane moduli in the flux induced superpotential or D-brane instanton induced superpotential corrections. We will now discuss these in turn.

Let us generally denote the brane deformation moduli as $\zeta^A$.\footnote{This is to distinguish the general brane deformation moduli from the particular combination of moduli called $y$ in the previous part of this article that is a linear combination of the moduli for the two branes ${\cal D}_a$ and ${\cal D}_b$.}
The K\"ahler potential for the axio-dilaton receives $\zeta^A$-dependent corrections, computed by dimensional reduction in \cite{Jockers:2004yj,Jockers:2005zy} schematically as
\bea
\label{Kahlercorr}
K = - {\rm log} \left(-i(S - \ov S)  + i {\cal L}_{A \ov B}  \zeta^A \ov \zeta^{\ov B}\right) + \ldots
\eea
for some coupling matrix ${\cal L}_{A \ov B}$. Here $S$ denotes the ${\cal N}=1$ chiral superfield related to the combination $\tau = C_0 + \frac{i}{g_s}$ by a $\zeta^A$-dependent shift, see \cite{Jockers:2004yj,Jockers:2005zy}  for details, and \cite{Grimm:2011dx,Kerstan:2011dy} for a recent discussion in the mirror symmetric context.
These corrections are universal in that they cannot be avoided by a particular choice of geometric setup or fluxes.\footnote{This is in contrast to a similar correction of the K\"ahler potential for the K\"ahler moduli, likewise discussed in \cite{Jockers:2004yj,Jockers:2005zy}, which can in principle be avoided for branes along 4-cycles $\Sigma$ that do not couple to the Ramond-Ramond field moduli $c^a - \tau b^a$, obtained by dimensional reduction of $C_2 - \tau B_2$ along elements of $H^{1,1}_-(\Sigma)$. }
Even in absence of an explicit dependence of the superpotential on the brane moduli responsible for inflation, the stabilisation of the axio-dilaton $S$ by 3-form background fluxes  in the spirit of \cite{Giddings:2001yu} will therefore induce a $\zeta^A$-dependence of the scalar potential due to the appearance of the K\"ahler potential in the K\"ahler covariant derivative $D_S W = \partial_S W + K_{,S} \, W$. This effect is analogous to the entanglement of K\"ahler modulus stabilisation with a mass term for D3-brane moduli in the scenario envisaged in \cite{Kachru:2003sx}.
It turns out that the resulting mass term for $\zeta^A$ is proportional to $g_s |W_0|^2$, where $W_0$ is the value of the superpotential in the minimum. 
This mass term alone can therefore in principle be avoided by stabilisation in the extreme perturbative regime $g_s \ll1$ or by $W_0 \ll 1$.
These options, however, are constrained via the general link of $F$- and $D$-terms in supergravity (see e.g.\ \cite{Villadoro:2005yq} for a recent discussion) with the $D$-term uplift in the K\"ahler moduli sector pointed out above. In addition, it must be ensured that the mass of the remaining moduli stays sufficiently high in this regime. Both these issues depend on the particular stabilisation mechanism and require further investigation \cite{wip}.

The 7-brane moduli are known to appear also explicitly in the flux-induced superpotential.
Recall that the Type IIB flux superpotential can be written as \cite{Witten:1992fb,Aganagic:2000gs,Lust:2005bd,Jockers:2005zy,Martucci:2006ij}
\bea
\label{super}
W = \int_{X_3} \Omega \wedge (F_3 - \tau H_3) + \int_{{\cal C}_5} \Omega \wedge F.
\eea
Here $F_3$ and $H_3$ denote the Ramond-Ramond and Neveu-Schwarz three-form flux entering the Gukov-Vafa-Witten (GVW) superpotential  \cite{Gukov:1999ya}. The (half-)integer quantised gauge flux $F$ along the 7-brane divisor $\Sigma$ is continued to a 5-chain ${\cal C}_5$ ending on $\Sigma$.

Let us first consider the second term induced by gauge flux $F \in H^2(\Sigma)$. Note that  $H^2(\Sigma)$ splits into two-forms that are the pullback of elements in $H^2(X_3)$ and those which cannot be written as a pullback from the ambient space. The latter are elements of $H^{2}_{var}(\Sigma)$, the second relative cohomology \cite{Lerche:2002yw} of the divisor $\Sigma$ within the Calabi-Yau space $X_3$.
The important point is that only 
gauge flux $F$ with values in $H^{2}_{var}(\Sigma)$ leads to a direct $F$-term \cite{Lust:2005bd,Jockers:2005zy} via the second term in (\ref{super}) stabilising the brane deformation moduli.\footnote{More generally, see  \cite{Jockers:2008pe,Grimm:2008dq,Alim:2009rf,Alim:2009bx,Jockers:2009mn,Grimm:2009ef,Aganagic:2009jq,Grimm:2009sy,Jockers:2009ti,Grimm:2010gk}   for recent progress in the computation of brane superpotentials using various techniques.} 
The moduli are stabilised such that the gauge flux be of Hodge type $(1,1)$ along $\Sigma$. This enforces the 8-dimensional $F$-term supersymmetry condition \cite{Marino:1999af}
\bea
\label{MMMS}
{\cal F} = 2 \pi \alpha' F + B \in H^{(1,1)}(\Sigma).
\eea
Conversely, if we focus solely on gauge flux that does descend from two-forms in the ambient geometry, this type of superpotential is absent. The reason is that on genuine Calabi-Yau spaces with full $SU(3)$ holonomy $H^{(2)}(X_3)= H^{(1,1)}(X_3)$ and gauge flux $F$ in the pullback of $H^{(1,1)}(X_3)$ automatically satisfies (\ref{MMMS}). 
 Unlike the previous $\zeta^A$-dependent corrections in the K\"ahler potential, the second term in (\ref{super}) can therefore be avoided by restricting to gauge flux descending form the ambient space. Note that such fluxes  are precisely the ones that enter the $D$-term and thus generate the inflaton potential.

In addition, however, a more detailed evaluation of the periods entering the GVW-superpotential  $W = \int_{X_3} \Omega \wedge (F_3 - \tau H_3) $ in \cite{Jockers:2009ti} suggests that 
even in absence of gauge flux the brane deformation moduli can appear explicitly in the superpotential. This is also in agreement with the analysis of \cite{Gorlich:2004qm,Lust:2005bd, Denef:2005mm} for the special case of F-theory on $K3 \times K3$.
Whether or not a suitable choice of background 3-form fluxes $F_3$ and $H_3$ can avoid the appearance of such an $F$-term in a setup that still stabilises all complex structure moduli and the axio-dilaton is an interesting yet complicated question that requires knowledge of the detailed form of the periods in a concrete example.

Another, but possibly related effect was observed in \cite{Gomis:2005wc}:  The choice of 3-form flux $H_3$, required to stabilise $\tau$ via the GVW-superpotential, implies a non-trivial non-closed $B$-field such that $H_3 = d B$. This $B$-field can likewise stabilise some of the brane moduli by the $F$-term supersymmetry condition \cite{Marino:1999af}. This happens whenever the pullback of $B$ to the brane divisor is of type $(2,0)$. Again the supersymmetry condition (\ref{MMMS}) fixes some of the brane deformations. 
Conversely, the stabilisation of a certain 7-brane modulus via the effect of \cite{Gomis:2005wc} can be avoided by turning on appropriate $H_3$-flux and underlying $B$-field.
The challenge is to find $H_3$-flux generic enough to stabilise on the one hand the axio-dilaton $\tau$ and, together with $F_3$, all complex structure moduli while leaving at the same time the envisaged inflaton massless. We demonstrate this important point in App.~\ref{app_moduli} by constructing a simple toroidal example, which is a straightforward generalisation of  the setup in \cite{Gomis:2005wc}, where all complex structure moduli and the dilaton are fixed in a supersymmetric Minkowski vacuum without fixing all 7-brane moduli. Even on genuine Calabi-Yau spaces it is to be expected that stabilisation of $\tau$ generically leaves enough freedom to choose $H_3$ and the $B$-field without fixing all 7-brane moduli.

Finally, D3-brane instantons or gaugino condensation on D7-branes are likely to yield an explicit $\zeta^A$-moduli dependence of the superpotential via loop-corrections  in the gauge kinetic function appearing in the instanton partition function that depends on the flux induced D3-brane charge on the moving D-brane. This effect would be the direct analogue of the superpotential for D3-brane moduli studied in \cite{Berg:2004ek,Baumann:2006th} and was suggested from a different perspective in the context F-theory model building in \cite{Marchesano:2009rz}.

To summarise, in concrete examples a detailed analysis of the $F$-term superpotential for the brane deformation moduli is required. 
Two possible approaches are conceivable. First, one may try to forbid each single $F$-term by itself without tuning order one expressions against each other.
As indicated above, the explicit inflaton dependence of the superpotential may indeed be avoidable by a suitable choice of background geometry, gauge and 3-form fluxes. By contrast, a significant appearance of the brane moduli in the scalar $F$-term potential induced by (\ref{Kahlercorr}) can at best be circumvented by stabilising all moduli in a regime where $g_s \ll 1$ or $W_0 \ll1 $, which, however, is subject to strong extra constraints \cite{wip}. A probably more realistic, but less ambitious option is thus to invoke a moderate tuning in parameter space to achieve the mutual cancellation of the various perturbative and non-perturbative $F$-terms against each other. This is essentially the philosophy underlying D3 inflation models of the type \cite{Kachru:2003sx,Baumann:2007ah}. It remains to be seen if the need for such a tuning can be overcome in D7-brane inflation
 by the above mechanisms.

\subsection*{Acknowledgements}

We thank Ralph Blumenhagen, Andr\'{e}s Collinucci, Moritz K\"untzler, Christoph Mayrhofer and Christian Pehle for helpful comments. We are particularly grateful to Thomas Grimm, Hans Jockers and Peter Mayr for important discussions and correspondence. S.C.K.\ acknowledges support by the Studienstiftung des deutschen Volkes, St.St. thanks  the Cusanuswerk for support. A.H.\ and T.W.\ are grateful to the Max-Planck-Institute, Munich for hospitality. D.L.\ would like to thank the Theory Division of CERN for its hospitality. This work was supported by the Transregio TR33 "The Dark Universe" and partially by the Cluster of Excellence "Origin and Structure of the Universe" in Munich.

\appendix

\section{Supergravity computation of D7/D7 potential}\label{sugra}

\subsection{Derivation of supergravity solution for a magnetised D7-brane}
\label{app-derivation}
In this appendix we fill in some gaps of the discussion in Sect.~\ref{probe-approx} which was aimed at calculating the solution for a magnetised D7-brane in supergravity with fluxes given by (\ref{back-dense}). To obtain this solution we invoke the (extended) Buscher rules of T-duality on the supergravity background solution for a stack of $N$ D5-branes which are rotated by angles $\phi_b^1$ on $T^2_1$ and $\phi_b^2$ on $T^2_2$.
\par

We start from the expressions (\ref{D5metric}), (\ref{RRDilaton}), and (\ref{D5harmonic}). The barred coordinates are related to the unbarred ones by a simple rotation
\begin{equation}
\label{rotation}
 \begin{pmatrix}
  \bar{x}_4 \\ \bar{x}_5
 \end{pmatrix}
=
 \begin{pmatrix}
  \cos\phi_b^1 & \sin\phi_b^1 \\ -\sin\phi_b^1 & \cos\phi_b^1
 \end{pmatrix}
 \begin{pmatrix}
  x_4 \\ x_5
 \end{pmatrix}\,,
\end{equation}
and equivalently for the second torus. Writing the metric and $C_6$ in terms of the torus coordinates (without bars) we arrive at
\begin{equation}  
\begin{split}
ds^2
=&\,Z_5^{-\frac{1}{2}}\left(-dx_0^2+\ldots+ dx_3^2\right)+Z_5^{\frac{1}{2}}\left(dx_8^2+dx_9^2\right)\\
&+ Z_5^{-\frac{1}{2}}\left(\cos^2\phi_b^1 +\sin^2\phi_b^1 Z_5 \right)dx_4^2
+ Z_5^{\frac{1}{2}}\left(\cos^2\phi_b^1 +\sin^2\phi_b^1 Z_5^{-1}\right)dx_5^2\\
&+ Z_5^{-\frac{1}{2}}\left(\cos^2\phi_b^2 +\sin^2\phi_b^2 Z_5 \right)dx_6^2
+ Z_5^{\frac{1}{2}}\left(\cos^2\phi_b^2 +\sin^2\phi_b^2 Z_5^{-1}\right)dx_7^2\\
&+\sin\phi_b^1\cos\phi_b^1 \left(Z_5^{-\frac{1}{2}}-Z_5^{\frac{1}{2}}\right)\left(dx_4 dx_5 + dx_5 dx_4\right)\\
&+\sin\phi_b^2\cos\phi_b^2 \left(Z_5^{-\frac{1}{2}}-Z_5^{\frac{1}{2}}\right)\left(dx_6 dx_7 + dx_7 dx_6\right) ,
\end{split}\label{rotated_D5_solution}
\end{equation}
while the R-R-form potential $C_6$ is given by
\begin{equation}
\begin{alignedat}{2}
C_{012346}&=g_s^{-1}\cos\phi_b^1 \cos\phi_b^2 \left(Z_5^{-1}-1\right),\;
&C_{012347}&=g_s^{-1}\cos\phi_b^1 \sin\phi_b^2 \left(Z_5^{-1}-1\right),\\
C_{012356}&=g_s^{-1}\sin\phi_b^1 \cos\phi_b^2 \left(Z_5^{-1}-1\right),\;
&C_{012357}&=g_s^{-1}\sin\phi_b^1 \sin\phi_b^2 \left(Z_5^{-1}-1\right).
\end{alignedat}
\end{equation}

Now we perform T-duality using the Buscher rules \cite{Buscher:1987qj}: Taking $z$ to be the Killing coordinate along which the T-duality is performed the T-dual metric $\tilde{g}$ is given by
\begin{equation}
\tilde{g}_{\mu\nu}=g_{\mu\nu}-\left(g_{\mu z} g_{\nu z} - B_{\mu z} B_{\nu z}\right)/g_{zz}\,,\quad \tilde{g}_{\mu z} = B_{\mu z}/g_{zz}\,,\quad \tilde{g}_{z z} = 1 /g_{zz} \,,
\end{equation}
while the dilaton and the $B$-field transform to
\begin{equation}
\tilde{\phi}=\phi - \frac{1}{2}\log|g_{zz}|\,,\quad \tilde{B}_{\mu z} = g_{\mu z}/g_{zz}\,,\quad \tilde{B}_{\mu \nu} = B_{\mu\nu} + 2 g_{[\mu|z}B_{\nu]z}/g_{zz}\,.
\end{equation}
The extended Buscher rules for the R-R-forms are \cite{Bergshoeff:1995as,Meessen:1998qm,Myers:1999ps}
\begin{equation}
\begin{aligned}
\tilde{C}^{(n)}_{\mu_1\ldots \mu_{n}}&= C^{(n+1)}_{\mu_1\ldots \mu_{n} z}+n C^{(n-1)}_{[\mu_1\ldots\mu_{n-1}}B_{\mu_n]z}\\
&\quad+n(n-1)C^{(n-1)}_{[\mu_1\ldots\mu_{n-2}|z}  B_{|\mu_{n-1}|z} g_{|\mu_n]z}/g_{zz}\,,\\
\tilde{C}^{(n)}_{\mu_1\ldots \mu_{n-1}z}&= C^{(n-1)}_{\mu_1\ldots \mu_{n-1}}-(n-1) C^{(n-1)}_{[\mu_1\ldots \mu_{n-2}|z}g_{|\mu_{n-1}]z}/g_{zz}\,.
\end{aligned}
\end{equation}

To perform T-duality of the given solution of a rotated black extremal 5-brane along the directions $x_5$ and $x_7$ we first need those to be isometries. This can be achieved by the standard procedure of ``smearing'' along these directions (we closely follow \cite{Peet:2000hn}). Noting that the transverse Laplace equation is linear we can distribute an array of stacks of $N$ D5-branes  at distances $2\pi R n$ in the $x_5$ direction. This corresponds to the case of one single stack of $N$ D5-branes after compactifying the $x_5$ direction via $x_5 \sim x_5 + 2\pi R$. Taking $p$ general for a moment  we get (using \eqref{rotation})
\begin{align}
 \tilde{Z}_p = 1 + \sum_{n\in \mathbb{Z}}\frac{c_p g_s N {\alpha'}^{\frac{7-p}{2}}}{\left[x_\perp^2 +\left(\cos\phi_b^1\left(x_5-2\pi R n\right)-\sin\phi_b^1 x_4\right)^2\right]^{\frac{7-p}{2}}}.
\end{align}
Now for $2\pi R \ll x_\perp$, a tiny compact circular dimension of radius $R$, we can approximate the sum by an integral and obtain
\begin{align}
 \tilde{Z}_p &= 1 + c_p g_s N {\alpha'}^{\frac{7-p}{2}}\frac{1}{2\pi R \cos\phi_b^1} \int du \frac{1}{\left[x_\perp^2 +u^2\right]^{\frac{7-p}{2}}}\\
 &=1+ c_{p+1} g_s^\prime N \frac{1}{\cos\phi_b^1} \left(\frac{\sqrt{\alpha'}}{x_\perp}\right)^{7-(p+1)} ,
\end{align}
where $g_s^\prime = g_s \frac{\sqrt{\alpha^\prime}}{R}$. As $x_5$ is now an isometry of the SUGRA solution we can regard the solution \eqref{rotated_D5_solution} with $Z_5$ replaced by $\tilde{Z}_5$ as compactified on a circle along $x_5$ with infinitesimal radius $R$ and use the given T-duality rules to uncompactify this dimension.

If we now repeat this procedure along $x_7$, we see that the integral $\int du \frac{1}{\sqrt{r^2 + u^2}}$ diverges ($(\ell_s r)^2= x_8^2 + x_9^2$). We can however introduce an IR cutoff $R$ and thus regularise the integral by
\begin{equation}
 \int_0^c du \frac{1}{\sqrt{r^2 + u^2}} = -\log r + \log\left(c+\sqrt{r^2 + c^2}\right) \approx -\log\left(\frac{r}{R}\right).
\end{equation}
This cutoff also appears in the string calculation and the potential arising from it can be interpreted as a tadpole. Furthermore, it corresponds to the length scale at which the dilaton is normalised to $e^\phi = g_s$.

Equipped with the given T-duality rules one can deduce the new metric for a D7-brane with flux on two tori. Since we originally had a vanishing NS-NS field $B_{\mu\nu}$ in the T-duality directions we do not get any off-diagonal terms in our new metric. The metric then reads
\begin{equation}
d\tilde{s}^2=Z_7^{-\frac{1}{2}}ds^2(\mathbb{E}^{1,3})+Z_7^{-\frac{1}{2}}H_1 ds^2(\mathbb{E}^{2}_{45})+Z_7^{-\frac{1}{2}}H_2 ds^2(\mathbb{E}^{2}_{67})+Z_7^{\frac{1}{2}}ds^2(\mathbb{E}^{2}_{89})\,,
\end{equation}
where we have introduced the abbreviation $H_{1/2}=\left(\cos^2\phi_b^{1/2} + \sin^2\phi_b^{1/2} Z_7^{-1}\right)^{-1}$. Note that, as just explained, our $Z_7$ is given by
\begin{equation}
Z_7 = 1 - N \frac{g_s}{2\pi} \frac{1}{\cos\phi_b^1\cos\phi_b^2}\log{\frac{r}{R}}. 
\end{equation}
The additional $(\cos\phi_b^i)^{-1}$ factors explain the small mismatch of the potentials in \cite{Dasgupta:2002ew} (which of course vanishes for small $\phi^i$) where first the D7-brane probes the D3 background and then the other way around. For the new Kalb-Ramond $B$-field the only non-vanishing terms one finds are
\begin{align}
\tilde{B}_{45} = -\tan\phi_b^1 + \tan\phi_b^1 Z_7^{-1} H_1\quad\text{and}\quad \tilde{B}_{67} =  -\tan\phi_b^2 + \tan\phi_b^2 Z_7^{-1} H_2 .
\end{align}
The dilaton is easily calculated to be 
\begin{equation}
 e^{2\tilde{\phi}}= g_s^2 Z_7^{-2} H_1 H_2 ,
\end{equation}
while in general we get non-vanishing $C_4$, $C_6$ and $C_8$ fields indicating the existence of dissolved lower dimensional branes in the world-volume of the D7-brane \cite{Douglas:1995bn}:
\begin{align}
\label{C-field}
\tilde{C}_{0123} &= g_s^{-1}\left(Z_7^{-1}-1\right)\sin\phi_b^1\sin\phi_b^2,\nonumber \\ 
\tilde{C}_{012345} &=g_s^{-1}\left(Z_7^{-1}-1\right)\cos\phi_b^1\sin\phi_b^2 H_1,\nonumber\\
\tilde{C}_{012367} &=g_s^{-1}\left(Z_7^{-1}-1\right)\sin\phi_b^1\cos\phi_b^2 H_2,\\
\tilde{C}_{01234567} &=g_s^{-1}\left(Z_7^{-1}-1\right)\cos\phi_b^1\cos\phi_b^2 H_1 H_2 .\nonumber
\end{align}

By looking at some limiting cases we can check that these potentials are the ones to be expected. For $\phi_b^1=\phi_b^2=0$ we T-dualise perpendicular to the unrotated D5-branes and get a stack of normal D7-branes. Thus only $\tilde{C}_8$ is non-vanishing and the dilaton has the standard dependence $e^{2\phi}= g_s^2 Z_7^{\frac{3-p}{2}}$. For $\phi_b^1=\phi_b^2=\frac{\pi}{2}$ we T-dualise along the brane twice and get the background of a smeared stack of standard D3-branes; an unsmeared stack would of course have $Z_7\rightarrow Z_3$. The dilaton is constant as expected and only $C_4$ is non-vanishing (neglecting the additional subtlety of the self-duality constraint on $F_5$).\\

\subsection{(Anti-)Self-dual flux decomposition}\label{SD_ASD_F}
We now interpret the flux-dependent prefactor of the distance-dependent term in our potential in terms of (anti-)self-dual flux. To this aim we take the background of an extremal D7-brane without flux, which thus only sources the RR-field $C_8$. We then put a probe D7-brane with some arbitrary gauge flux $F$ into this background and calculate its potential.

The background is (again) given by \eqref{D7-flux-background1} and following expressions (this time however with $\phi_b^i=0$), i.e. a relative warping of the parallel and orthogonal directions of the brane
with dilaton $e^{2\phi}= g_s^2 Z_7^{-2}$ and RR potential $C_{01234567}=g_s^{-1}\left(Z_7^{-1}-1\right)$, where the distance dependence comes from the warp-factor $Z_7\sim \log r$. Using this background we can take the DBI-action
\begin{equation}
\begin{split}
  S_{\operatorname{DBI}} &=-T_7 \int d^8 \sigma \,e^{-\phi} \sqrt{-\det\left(g_{MN}+\mathcal{F}_{MN}\right)} \\
   &= -T_7 \int d^8 \sigma \,e^{-\phi} \sqrt{-g}\,\exp\left(\frac{1}{2} \operatorname{Tr} \log \left(\delta^M_{\;N} + \mathcal{F}^M_{\;N}\right) \right)
\end{split}
\label{DBI_again}
\end{equation}
and expand the last exponential in terms of the components of the flux $\mathcal{F}$, which up to order $\mathcal{O}\left(|\mathcal{F}|^6\right)$ gives
\begin{equation*}
 1+ \frac{1}{4} \mathcal{F}_{MN}\mathcal{F}^{MN}-\frac{1}{8}\mathcal{F}^A_{\;B}\mathcal{F}^B_{\;C}\mathcal{F}^C_{\;D}\mathcal{F}^D_{\;A}+\frac{1}{2}\left(\frac{1}{4}\mathcal{F}_{MN}\mathcal{F}^{MN}\right)^2.
\end{equation*}
Now, neglecting the kinetic terms, we have $-T_7 \int d^8 \sigma e^{-\phi} \sqrt{-g}=-2\pi\ell_s^{-4} \int d^4x\; \mathcal{V}_\parallel Z_7^{-1}g_s^{-1} $. So when we put together again \eqref{DBI_again}, the first term is $\sim Z_7^{-1}$. This divergence however is cancelled by part of the CS contribution $ S_{\operatorname{CS}}=\mu_7 \int C_8$. The second term $\mathcal{F}_{MN}\mathcal{F}^{MN}\propto Z_7$ enjoys the aforementioned generalised conformal symmetry. So, the only relevant part which is $r$-dependent is the quartic term. This however simplifies
\begin{align}
 -\frac{1}{8}\mathcal{F}^A_{\;B}\mathcal{F}^B_{\;C}\mathcal{F}^C_{\;D}\mathcal{F}^D_{\;A}+\frac{1}{2}\left(\frac{1}{4}\mathcal{F}_{MN}\mathcal{F}^{MN}\right)^2=-\frac{1}{8}\mathcal{F}^{+}_{AB}\mathcal{F}^{+ AB}\mathcal{F}^{-}_{CD}\mathcal{F}^{-CD}
\end{align}
upon defining the self-dual and anti-self-dual parts of the field strength as
\begin{equation}
\begin{split}
 \mathcal{F}^{+}_{AB} & = \frac{1}{2}\left( \mathcal{F}_{AB}+\frac{1}{2}\epsilon_{AB}^{\quad CD} \mathcal{F}_{CD} \right) = * \mathcal{F}^+_{AB}\,,\\
\mathcal{F}^{-}_{AB} & = \frac{1}{2}\left(\mathcal{F}_{AB}-\frac{1}{2}\epsilon_{AB}^{\quad CD} \mathcal{F}_{CD} \right)= - * \mathcal{F}^-_{AB}\,.
\end{split}
\end{equation}
This term goes as $\sim Z_7^{-1+2}$ and thus yields our logarithmic potential. Thus it is the \mbox{(anti-)self-dual} parts of the relative flux which govern the strength of the interaction potential in the case of toroidal compactification. For a Calabi-Yau orientifold this very result has been generalised within Sect.~\ref{Pot-gen} or more specifically in eq.~\eqref{CY_potential}.

\section{Evaluating the one-loop integral}\label{app_1loop}
In this appendix we provide more details for the calculation of the inter-brane potential via an open string one-loop amplitude in Sect.~\ref{1loop}. In particular, we will evaluate the modular functions in the integrand of (\ref{loop-amplitude}) and arrive at the crucial result
\begin{equation}\label{integrand}
 -i \sum_{\alpha, \beta \in \{0,1/2\}}\eta_{\alpha\beta} e^{i\pi\delta_{ab}(1-2\beta)} \frac{\vartheta \myfrac{\alpha}{\beta}(0,it)^3}{\eta(it)^9}\frac{\vartheta\myfrac{\alpha+\delta_{ab}}{\beta}(0,it)}{\vartheta\myfrac{1/2+\delta_{ab}}{1/2}(0,it)} \approx (t\phi_{ab})^3 
\end{equation}
for $|t\phi_{ab}|\ll 1$. This allows us to write the integral in the simple form (\ref{approx-integral}). As in the rest of this paper, we restrict ourselves to the case of small flux, i.e.\ $|\phi_{ab}| \ll 1$. Furthermore, without loss of generality we assume $\phi_{ab}>0$.
\par

In order to approximate the integrand of (\ref{loop-amplitude}) we distinguish between three different regimes: $t \ll 1$, $1\ll t\ll 1/\phi_{ab}$, and $1/\phi_{ab}\ll t$.
\par

We first consider the regime in which $t\ll 1$. Using the modular transformation properties (\ref{TMod1}) and (\ref{TMod2}) we find
\begin{align}
 -i \sum_{\alpha, \beta \in \{0,1/2\}}\eta_{\alpha\beta} &  e^{i\pi\delta_{ab}(1-2\beta)} \frac{\vartheta \myfrac{\alpha}{\beta}(0,it)^3}{\eta(it)^9}\frac{\vartheta\myfrac{\alpha+\delta_{ab}}{\beta}(0,it)}{\vartheta\myfrac{1/2+\delta_{ab}}{1/2}(0,it)}\nonumber \\
& = -t^3 \sum_{\alpha, \beta \in \{0,1/2\}}\eta_{\alpha\beta} \frac{\vartheta \myfrac{-\beta}{\alpha}(0,i/t)^3}{\eta(i/t)^9}\frac{\vartheta\myfrac{-\beta}{\alpha+\delta_{ab}}(0,i/t)}{\vartheta\myfrac{1/2}{1/2+\delta_{ab}}(0,i/t)} .
\end{align}
Using Riemann's identity (\ref{Riemann}) this expression can be further simplified to give
\begin{equation}
 -t^3 \sum_{\alpha, \beta \in \{0,1/2\}}\eta_{\alpha\beta} \frac{\vartheta \myfrac{-\beta}{\alpha}(0,i/t)^3}{\eta(i/t)^9}\frac{\vartheta\myfrac{-\beta}{\alpha+\delta_{ab}}(0,i/t)}{\vartheta\myfrac{1/2}{1/2+\delta_{ab}}(0,i/t)} = - 2 t^3 \frac{\vartheta \myfrac{1/2}{1/2 + \delta_{ab}/2}(0,i/t)^4}{\eta(i/t)^9\vartheta\myfrac{1/2}{1/2+\delta_{ab}}(0,i/t)} .
\end{equation}
The modular functions have a simple expansion in $q = \exp(2\pi i \tau)$, namely (\ref{expEta}) and (\ref{expTheta}). In the present case, $\tau = i/t$ i.e.\ $q = \exp(-2\pi/t)$ which is a small number if $t\ll 1$. Applying this expansion yields
\begin{equation}\label{result_withMod}
  - 2 t^3 \frac{\vartheta \myfrac{1/2}{1/2 + \delta_{ab}/2}(0,i/t)^4}{\eta(i/t)^9\vartheta\myfrac{1/2}{1/2+\delta_{ab}}(0,i/t)} \approx 2^4 t^3 \frac{\sin(\phi_{ab}/2)^4}{\sin(\phi_{ab})}    \approx (t\phi_{ab})^3 .
\end{equation}
\par

\begin{figure}[t]
 \centering
 \includegraphics[width = 0.7\textwidth]{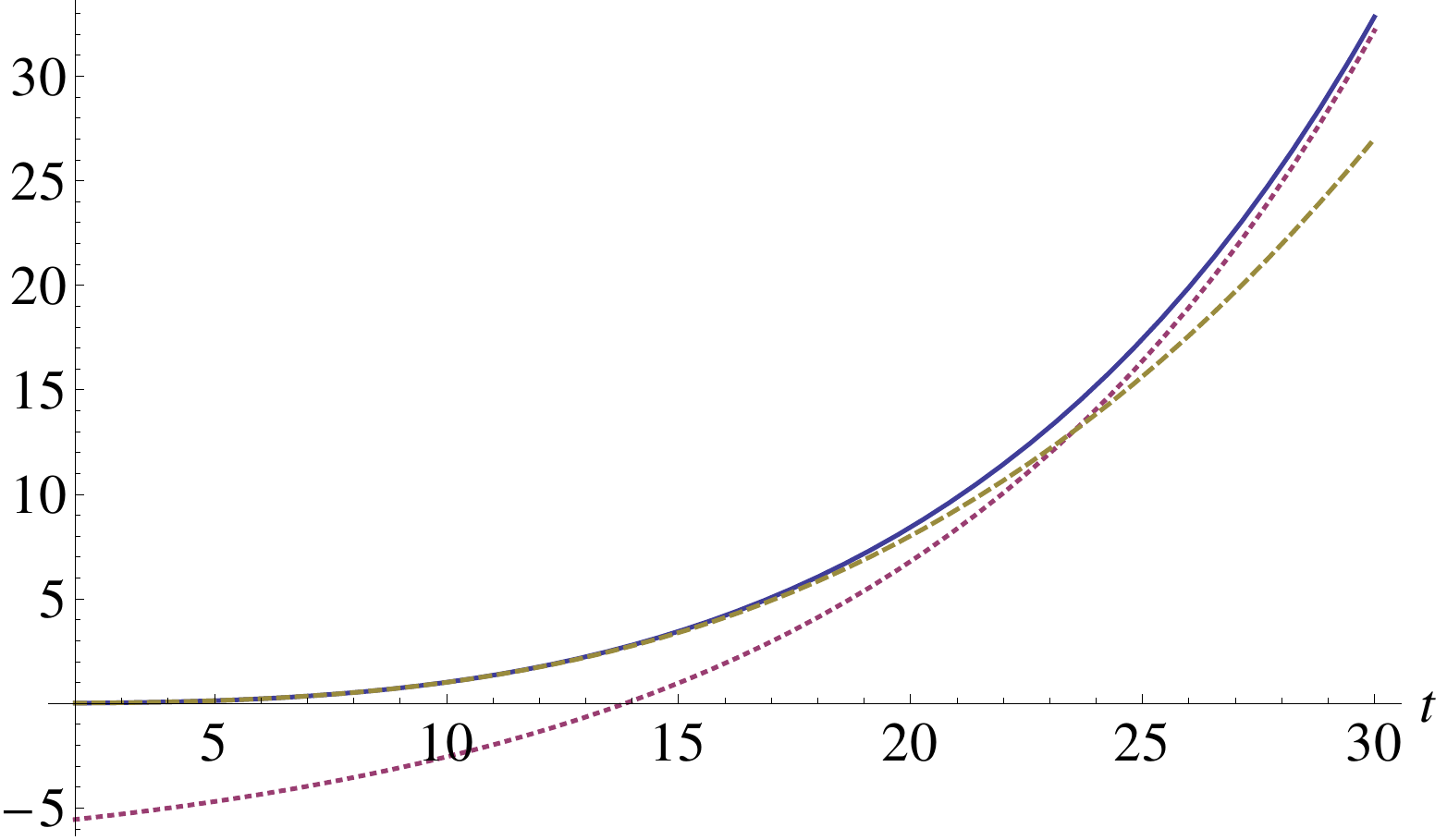}
 \caption{Asymptotic behaviour of the left-hand side of (\ref{integrand}) for $\phi_{ab} = 0.1$. The graph shows the full expression (solid) as well as the approximations for $0\ll t\ll 1/\phi_{ab}$ (dashed) and for $t\gg 1/\phi_{ab}$ (dotted).}
 \label{fig:limitModFunc}
\end{figure}
Now we move to the regime in which $1\ll t\ll 1/\phi_{ab}$. Using
\begin{equation}
 \vartheta\myfrac{\alpha + \delta}{\beta}(0,it) = q^{\frac{\delta^2}{2}}e^{2\pi i \beta \delta} \vartheta \myfrac{\alpha}{\beta}(i t \delta, it)
\end{equation}
as well as Riemann's identity the expression on the left-hand side of (\ref{integrand}) becomes
\begin{align}
 -i \sum_{\alpha, \beta \in \{0,1/2\}}\eta_{\alpha\beta} & e^{i\pi\delta_{ab}(1-2\beta)} \frac{\vartheta \myfrac{\alpha}{\beta}(0,it)^3}{\eta(it)^9}\frac{\vartheta\myfrac{\alpha+\delta_{ab}}{\beta}(0,it)}{\vartheta\myfrac{1/2+\delta_{ab}}{1/2}(0,it)} \nonumber\\
&= -2i \frac{\vartheta \myfrac{1/2}{1/2}\left(\frac{i t \delta_{ab}}{2},it\right)^4}{\eta(it)^9 \vartheta\myfrac{1/2}{1/2}(i t \delta_{ab},it)}  .
\end{align}
In this case, $q = \exp(-2\pi t)$ which is small for $t\gg 1$. Thus, the expansion of the modular functions in $q$ is again applicable. We find
\begin{align}
 \vartheta \myfrac{1/2}{1/2} (i t \delta_{ab}, it) & = - iq^{\frac{1}{8}} \sum_{n=0}^{\infty} (-1)^n q^{\frac{n(n+1)}{2}} \left\{q^{-\delta_{ab}\left(n + \frac{1}{2}\right)} - q^{\delta_{ab}\left(n + \frac{1}{2}\right)}\right\} \nonumber\\
 & \approx - i q^{\frac{1}{8}} \left\{ q^{- \frac{\delta_{ab}}{2}} - q^{\frac{\delta_{ab}}{2}} \right\} . \label{q-expansion}
\end{align}
Expanding the result for small $t\phi_{ab}$ yields
\begin{equation}\label{result_noMod}
 -2i \frac{\vartheta \myfrac{1/2}{1/2}\left(\frac{i t \delta_{ab}}{2},it\right)^4}{\eta(it)^9 \vartheta\myfrac{1/2}{1/2}(i t \delta_{ab},it)} \approx 2 \frac{\left( q^{-\frac{\delta_{ab}}{4}} - q^{\frac{\delta_{ab}}{4}} \right)^4}{q^{-\frac{\delta_{ab}}{2}} - q^{\frac{\delta_{ab}}{2}} } \approx ( t \phi_{ab})^3 .
\end{equation}
Thus, at leading order (\ref{result_noMod}) exactly matches the result (\ref{result_withMod}) found in the $t\ll 1$ region.
\par

Finally, we consider the regime in which $t\gg 1/\phi_{ab}$. The treatment is very similar to the case where $1\ll t\ll 1/\phi_{ab}$. However, the expansion in $t\phi_{ab}$ is no longer applicable. Instead, we find
\begin{equation}\label{larget}
 -2i \frac{\vartheta \myfrac{1/2}{1/2}\left(\frac{i t \delta_{ab}}{2},it\right)^4}{\eta(it)^9 \vartheta\myfrac{1/2}{1/2}(i t \delta_{ab},it)} \approx 2 \left(q^{-\frac{\delta_{ab}}{2}} - 4\right) .
\end{equation}
\par

The situation is summarised in Fig.~\ref{fig:limitModFunc}. First of all we see that the expression on the left-hand side of (\ref{integrand}) is well approximated by $(t\phi_{ab})^3$ in the whole range $0\ll t\ll 1/\phi_{ab}$ and not only in $0\ll t\ll 1$ as one might naively expect. Furthermore, the figure illustrates that the asymptotic behaviour of the expression for $t\gg 1/\phi_{ab}$ is correctly given by (\ref{larget}). The relative error that one gets using the approximations (\ref{integrand}) and (\ref{larget}) instead of the full expression is small.
\par

Now we go back to (\ref{loop-amplitude}). As we have derived in Sect.~\ref{sec_geom}, the open string tachyon appears if the distance $r$ falls below the critical value $r_{\rm crit.}^2 = \phi_{ab}/(2\pi)$. Thus, during inflation, $r^2$ will be larger than $\phi_{ab}/(2\pi)$. At $t\gg 1/\phi_{ab}$ we have two competing effects: First of all, in view of (\ref{larget}), the expression on the left-hand side of (\ref{integrand}) grows exponentially as $\exp(t\phi_{ab})$. Second, the distance-dependent exponential in the integrand of (\ref{loop-amplitude}) gives an exponential suppression $\exp(-2\pi t r^2)$. Therefore, (\ref{loop-amplitude}) is convergent as long as $r > r_{\rm crit.}$, due to the effective exponential decline of the whole integrand. For $r \gg r_{\rm crit.}$ this exponential effectively cuts off the integral in the region $t \gg 1/(2\pi r^2)$. In this case the one-loop integral has contributions only from regions where the left-hand side of (\ref{integrand}) is well approximated by $(t\phi_{ab})^3$. We conclude that as long as $r^2 \gg \phi_{ab}/(2\pi)$ the expression (\ref{loop-amplitude}) is well approximated by (\ref{approx-integral}).

\section{Modular functions}\label{app_modfunc}
In this appendix we collect the definitions and some useful formulae for the modular functions. An exhaustive treatment can be found in \cite{Mumford:Theta}. The modular functions are defined by 
\begin{align}
 \vartheta\myfrac{\alpha }{ \beta}(\nu,\tau) =& \sum_{n\in\mathbb{Z}}\exp\left[\pi i (n+\alpha)^2\tau + 2\pi i (n+\alpha) (\nu + \beta)\right] \nonumber\\
 =& e^{2\pi i \alpha  (\nu + \beta) }q^{\alpha^2/2}\nonumber\\
 & \times \prod_{n = 1}^{\infty}(1-q^n)(1+e^{2\pi i  (\nu + \beta)}q^{n-1/2 +\alpha})(1+e^{-2\pi i  (\nu + \beta)}q^{n-1/2 -\alpha}) ,\\
 \eta(\tau)=& q^{\frac{1}{24}}\prod_{n=1}^{\infty}(1-q^n) = q^{\frac{1}{24}} \sum_{n=-\infty}^{\infty}(-1)^n q^{\frac{3n^2 - n}{2}} \label{expEta} ,
\end{align}
where $q = \exp(2\pi i \tau)$. The function $\vartheta\myfrac{1/2}{1/2}(\nu,\tau)$ may be rewritten in a slightly more convenient form:
\begin{equation}\label{expTheta}
 \vartheta_1 (\nu,\tau) \equiv -\vartheta\myfrac{1/2}{1/2}(\nu,\tau) =  2 q^{\frac{1}{8}} \sum_{n=0}^{\infty} (-1)^n q^{\frac{n(n+1)}{2}} \sin\left(2\pi\nu\left(n + \frac{1}{2}\right)\right) .
\end{equation}
The theta functions enjoy the invariance
\begin{equation}
 \vartheta\myfrac{\alpha \pm 1}{\beta}(\nu,\tau) = \vartheta\myfrac{\alpha}{\beta}(\nu,\tau) .
\end{equation}
The following T-modular transformations are relevant in the main text:
\begin{align}
 \eta(it) &= t^{-1/2}\eta(i/t) , \label{TMod1}\\
 \vartheta\myfrac{\alpha }{ \beta}(0,it) &= t^{-1/2} e^{2\pi i \alpha \beta}\vartheta\myfrac{-\beta }{ \alpha}(0,i/t) . \label{TMod2}
\end{align}
Furthermore, we make use of Riemann's identity
\begin{equation}\label{Riemann}
 \frac{1}{2}\sum_{\alpha,\beta \in \{0,1/2\}}\eta_{\alpha\beta}\prod_{i=0}^4\vartheta \myfrac{\alpha}{\beta}(\nu_i, \tau) = \prod_{i=0}^4\vartheta \myfrac{1/2}{1/2}(\nu'_i, \tau) ,
\end{equation}
where
\begin{equation}
 \left(\begin{matrix}\nu'_1\\\nu'_2\\\nu'_3\\\nu'_4\end{matrix}\right) = \frac{1}{2}\left(\begin{matrix}1&1&1&1\\1&1&-1&-1\\1&-1&1&-1\\1&-1&-1&1\end{matrix}\right)\left(\begin{matrix}\nu_1\\\nu_2\\\nu_3\\\nu_4\end{matrix}\right) .
\end{equation}

\section{7-brane potential from {\boldmath ${\cal N}=1$} effective action}
\label{corr}

In this appendix we present an alternative derivation of the attractive potential of two magnetised 7-branes. We will motivate an interpretation of  the 7-brane  potential in terms of corrections to quantities that appear in an effective four-dimensional supersymmetric description of the model. This is  somewhat heuristic because we will take the limit of infinite transverse volume corresponding to the limit of rigid supersymmetry; more work is needed to make the proposal more rigorous 
in a fully-fledged super{\emph{gravity}} analysis.

In  field theory language the attractive potential between branes with non-trivial relative flux  results from a $D$-term
\begin{equation}\label{D-term}
 V_D = \frac{1}{2}\Re (f)^{-1} \xi^2,
\end{equation}
where $\xi$ is the (field-dependent) Fayet-Iliopoulos term and $f$ is the holomorphic gauge kinetic function.
Both quantities $\xi$ and $f$ receive corrections from one-loop diagrams.
In this appendix we distinguish between the tree-level and one-loop pieces as
\bea
 \left. \xi\right|_{\rm 1-loop} &= \xi_0 + \delta \xi, \quad\quad
 \left.\Re (f)\right|_{\rm 1-loop} &= \Re (f_0) + \Delta.
\eea
This gives a potential
\begin{equation}\label{expanded-potential}
 V \approx \frac{1}{2}\Re (f_0)^{-1}\, \xi_0^2 - \frac{1}{2}\Re (f_0)^{-2} \, \xi_0^2 \, \Delta + \Re (f_0)^{-1} \, \xi_0 \, \delta \xi.
\end{equation}
In principle these corrections can be calculated and the purpose of the present section is to show that they are of the right form to reproduce the logarithmic potential that attracts the branes towards each other.

For simplicity we will again work with a toroidal compactification in the limit of infinite volume of the torus where the branes are separated (i.e.\ $T^2_3$). 
Let us start with  non-zero flux  on $T^2_1$ only. Then, comparison with the standard DBI action expanded for small flux (cf.~(\ref{pot_const})) yields
\begin{align}
 \xi_0 =  \frac{2\pi\Re (f_0)}{\ell_s^2} \frac{\left(\phi_{a} - \phi_b \right)}{\sqrt{2}} \label{FI-term}, \quad\quad 
 \Re (f_0) = \frac{\mathcal{V}_{||}}{2\pi g_s}.
\end{align}

We first turn to corrections to $\xi$.
In a fully-fledged supergravity analysis, the field-dependent Fayet-Iliopoulos term is determined by the gauging of chiral multiplets under the associated abelian symmetry and depends on the K\"ahler potential. Corrections to $\xi$ are ultimately rooted in corrections to the K\"ahler potential.
In this appendix we content ourselves with the interpretation of $\xi$ as an albeit field-dependent $D$-term. Instead of thinking about one-loop corrections to the K\"ahler potential we directly compute one-loop corrections to the $D$-term in a string calculation which to the best of our understanding is really justified in the situation of rigid supersymmetry. 

In \cite{Atick1987, Dine1987} the generation of a Fayet-Iliopoulos term at the one-loop level was studied in heterotic string theory by calculating scalar masses which are generated by this term. For this purpose the authors identify a vertex operator for the auxiliary $D$-field and evaluate one-loop string diagrams with one insertion of this vertex operator. This approach is adapted to a system of intersecting D6-branes in Type IIA string theory in \cite{Lawrence2004b}, which is T-dual to the setup of magnetised D7-branes dealt with in the present paper. The authors of \cite{Lawrence2004b} are merely interested in the quadratic divergence of this correction which is present already if supersymmetry is preserved and which is cancelled as soon as global tadpole cancellation is taken into account. The treatment in \cite{Akerblom:2007uc} goes beyond this and calculates corrections to $\xi$ in the case where supersymmetry is broken at tree-level. The correction has the form
\begin{equation}
 \delta \xi_{ab} = -\frac{i}{2\pi \ell_s^2} \partial_{\nu}\int_0^{\infty} dt A_{ab}(\nu, it).
\end{equation}
$A_{ab}(\nu, it)$ is the partition function with one insertion of $\exp(2\pi i \nu J_0)$ evaluated on the annulus in the $(ab)$-sector. $J_0$ is the zero mode of the world-sheet $U(1)$ current which is identified with the vertex operator for the auxiliary $D$-field in the $(0)$-ghost picture \cite{Lawrence2004b, Akerblom:2007uc}.\par

For the case of non-vanishing flux  on only one torus (T-dual to D6-branes at  a non-vanishing relative angle $\phi_{ab}$ in one torus only) one finds
\begin{align}
 A_{ab} \propto& \mathcal{V}_2 I_{ab} \frac{1}{t^3}\frac{-i \exp\left(-2\pi t r^2\right) }{\eta(it)^9  \vartheta\myfrac{1/2}{1/2}(\delta_{ab} it, it)}\nonumber\\
& \times\vartheta\myfrac{1/2}{1/2}\left(\frac{3\nu}{2} + \frac{\delta_{ab}}{2} it, it\right)\vartheta\myfrac{1/2}{1/2}\left(-\frac{\nu}{2} + \frac{\delta_{ab}}{2} it, it\right)\vartheta\myfrac{1/2}{1/2}\left(-\frac{\nu}{2} - \frac{\delta_{ab}}{2} it, it\right)^2.
\end{align}
$\mathcal{V}_2$ is the volume along $T^2_2$ and $\delta_{ab} \equiv  (\phi_{a}-\phi_{b})/\pi \equiv \phi_{ab}/\pi$. $I_{ab}$ is the intersection number which for simplicity we assume to have the form $I_{ab} = p_a - p_b$ (i.e.\ the gauge theories living on the T-dual single-wrapped D7-branes are indeed just $U(1)$ theories \cite{Hashimoto:1997gm, Rabadan:2001mt}). Note that there will appear further constant factors in front of this amplitude depending on its normalisation. At this point we are a bit cavalier about these factors and simply observe that the result matches the one found in Sect.~\ref{1loop} parametrically. 
To this end we perform essentially the same steps that were needed to derive the potential in Sect.~\ref{1loop} and find
\begin{equation}
 \delta \xi_{ab} \propto \frac{ \mathcal{V}_2}{\ell_s^2}I_{ab} \frac{\cos(\phi_{ab}/2)\sin^3(\phi_{ab}/2)}{\sin(\phi_{ab})}\log\left(\frac{r}{R}\right).
\end{equation}
$I_{ab}$ can be replaced in terms of  $I_{ab} = p_a - p_b$ via (\ref{flux_quantisation_1}) to get
\begin{equation}
\label{FI-correctiona}
 \delta \xi_{ab} \propto \frac{\mathcal{V}_{||}}{\ell_s^2}  \frac{\cos(\phi_{ab}/2)\sin^3(\phi_{ab}/2)}{\cos\phi_a \cos\phi_b}\log\left(\frac{r}{R}\right).
\end{equation}
Inserted into (\ref{expanded-potential}) this yields the correct behaviour for small angles.\par

We now turn to the threshold corrections $\Delta$ (see also \cite{Dvali:2003zh}). Threshold corrections in string theory can be computed by means of the background field method \cite{Bachas:1992bh, Bachas:1996zt, Antoniadis:1999ge, Lust:2003ky}. However, recall that this method actually calculates threshold corrections to the physical gauge coupling $g_{\rm YM}$, which we will call $\Delta^g$. As is well-known, while at tree-level this coupling is related to the real part of the gauge kinetic function as $\Re(f) = 1/g_{\rm YM}^2$ they start to differ at one loop \cite{Kaplunovsky:1993rd}. There are corrections to $g_{\rm YM}$ which are holomorphic functions of the superfields and there are corrections that are not holomorphic. Only the holomorphic part of these corrections appear in $f$. To calculate the threshold corrections one evaluates the free energy $\bbF$ in the background of a spacetime magnetic field $\ccB$ at one-loop,
\begin{equation}
\bbF(\ccB)|_{\rm 1-loop} = \ccA(\ccB) + \mathcal{M}(\ccB) + \mathcal{K} + \mathcal{T},
\end{equation}
and deduces
\begin{equation}
 \frac{1}{g_{\rm YM}^2} = \frac{1}{(2\pi)^2}\frac{\partial^2}{\partial \ccB^2} \left(\frac{\bbF(\ccB)}{\mathcal{V}_{\rm 4d}}\right),
\end{equation}
where $\mathcal{V}_{\rm 4d}$ is the regularised volume of the non-compact dimensions.
\par

The calculation of the annulus amplitude of an open string stretched between two branes in the presence of a spacetime magnetic field was detailed in \cite{Bachas:1992bh, Bachas:1996zt, Antoniadis:1999ge, Lust:2003ky}.
In particular the gauged annulus amplitude in our situation reads
\begin{align}
 \ccA_{ab}(\ccB) = & -\frac{ \mathcal{V}_{\rm 4d}\mathcal{V}_{||}}{2^4} \ccB (\ccF^a_{45} - \ccF^b_{45}) \int_0^{\infty} \frac{dt}{t^3}\exp\left(-2\pi t r^2\right)\times\nonumber\\
&\times \sum_{\alpha, \beta \in \{0,1/2\}}\eta_{\alpha\beta} e^{i\pi(\delta_{ab}+ \epsilon)(1-2\beta)}
 \frac{\vartheta \myfrac{\alpha}{\beta}(0,it)^2}{\eta(it)^6}\frac{\vartheta\myfrac{\alpha + \epsilon}{\beta }(0,it)}{\vartheta\myfrac{1/2 + \epsilon}{1/2 }(0,it)} \frac{\vartheta\myfrac{\alpha+\delta_{ab}}{\beta}(0,it)}{\vartheta\myfrac{1/2+\delta_{ab}}{1/2}(0,it)},
\end{align}
where $\pi \epsilon = \arctan(\mathcal{B})$. Performing precisely the same steps as in App.~\ref{app_1loop} we arrive at
\begin{align}
 \ccA_{ab}(\ccB) =& -\mathcal{V}_{\rm 4d}\mathcal{V}_{||}\ccB (\ccF^a_{45} - \ccF^b_{45}) \frac{\sin\left(\frac{\pi \epsilon + \phi_{ab}}{2}\right)^2\sin\left(\frac{\pi \epsilon - \phi_{ab}}{2}\right)^2}{\sin(\pi\epsilon)\sin(\phi_{ab})} \ln\left(\frac{r}{R}\right)\nonumber \\
	 =& -\frac{ \mathcal{V}_{\rm 4d}\mathcal{V}_{||} }{4} F(\ccB)(\ccF^a_{45} - \ccF^b_{45})  \ln\left(\frac{r}{R}\right),
\end{align}
where
\begin{equation}
 F(\ccB) =  \ccB \frac{(\cos(\arctan(\ccB)) - \cos(\phi_{ab}))^2}{\sin(\arctan(\ccB))\sin(\phi_{ab})}.
\end{equation}
Noting that
\begin{equation}
 \left.\frac{\partial^2}{\partial \ccB^2}F(\ccB)\right|_{\ccB=0} = -\sin(\phi_{ab})
\end{equation}
we easily find a threshold correction
\begin{align}\label{total_gauge_correction}
 \Delta^g_{ab} = \frac{1}{(2\pi)^2} \left.\frac{\partial^2}{\partial \ccB^2 }  \left(\frac{\ccA_{ab}(\ccB)}{\mathcal{V}_{\rm 4d}}\right)\right|_{\ccB =0} = \frac{\mathcal{V}_{||} }{4 }\frac{1}{(2\pi)^2} \frac{\sin^2(\phi_{ab})}{\cos\phi_a \cos\phi_b} \ln\left(\frac{r}{R}\right).
\end{align}

However, this correction depends on the complex structure (Type IIA) or, respectively, the K\"ahler moduli (Type IIB) in a non-holomorphic manner. In particular, via the logic of \cite{Kaplunovsky:1993rd}, $\Delta^g$ does not furnish a holomorphic correction $\Delta$ to the gauge kinetic function $f$. Hence we do not include this correction in (\ref{expanded-potential}).\footnote{Curiously, 
for small relative angle $\phi_{ab}$ the form of the correction (\ref{total_gauge_correction}) fits  with the result for the potential obtained in Sect.~\ref{1loop} except for the sign. 
If we did include it in the potential  (\ref{expanded-potential}) it would combine with the correction (\ref{FI-correctiona}) and only modify the overall factor.
In any case we see that the corrections to $\xi$ are essential at this point to get the right answer.}

Let us briefly comment on the case of non-trivial flux on both $T^2_1$ and $T^2_2$.  Instead of going through the computation of $\delta \xi$ from scratch it is easier to adapt the following argument proposed in \cite{Akerblom:2007uc}: Consider a system of two fluxed branes with a small component of supersymmetry breaking flux. Then at leading order in $\xi$ there is a relation between $\delta \xi$ and $\Delta^g_{\rm SUSY}$ of the form
\begin{equation}
	\delta \xi \propto \xi_0 \, g_{\rm YM}^2 \, \Delta^g_{\rm SUSY}.
\end{equation}
Here $\Delta^g_{\rm SUSY}$ are the threshold corrections to $1/g_{\rm YM}^2$ evaluated after setting the non-super\-symmetric flux to zero. This in turn is just another confirmation of the fact that both corrections identified in (\ref{expanded-potential}) are of the same type. Application of the background field method proceeds as before and one finds that
\begin{equation}
\label{Delta-SUSY}
	\Delta^g_{\rm SUSY} \propto \mathcal{V}_{||}\frac{\sin \phi_{ab}^1 \sin \phi_{ab}^2}{\cos \phi_a^1 \cos \phi_b^1 \cos \phi_a^2 \cos \phi_b^2}\ln\left(\frac{r}{R}\right) .
\end{equation}
For a small supersymmetry breaking parameter $\xi$ this has the right form to reproduce (\ref{IIB-potential-1angle}).

\section{Moduli stabilisation in a  toroidal example}
\label{app_moduli}

Our aim is to demonstrate the existence of a Type IIB compactification with a 7-brane along a divisor $\Sigma$ and with background flux $H_3$ and $F_3$ satisfying the following properties: \\
i) The Gukov-Vafa-Witten superpotential stabilises all complex structure moduli and the dilaton in the perturbative regime. \\
ii) The $H_3$-flux satisfies the Freed-Witten anomaly $[H|_{\Sigma}] =0$, where $\Sigma$ denotes the divisor wrapped by a 7-brane. \\
iii) The $B$-field induced by the non-trivial $H_3$ does not stabilise all the 7-brane moduli.

In the context of toroidal orientifolds, such an example is easy to construct. Specifically we start from the setup introduced in Sect.~4.1. of \cite{Gomis:2005wc} (see also \cite{Camara:2004jj}), consisting of a Type IIB orientifold
on $T^2_1 \times T^2_2 \times T^2_3/ {\mathbb Z}_2 \times {\mathbb Z}_2$. Consider the 7-brane wrapping the first 2 tori $T^2_1 \times T^2_2$ with modulus $\zeta_3$ along the third torus $T^2_3$.
We switch on the following fluxes:
\bea
H_3 &=&  \ell_s^2 N \left( dx^1 \wedge dx^2 \wedge dx^3 + dy^1 \wedge dy^2 \wedge dx^3 \right) , \\
F_3 &=&  \ell_s^2 N \left( dx^1 \wedge dx^2 \wedge dy^3 + dy^1 \wedge dy^2 \wedge dy^3 + \Delta F_3 \right), \\
\Delta F_3 &=& i M \, dy^1 \wedge dy^2 \wedge dx^3 + M \, dx^1 \wedge dy^2 \wedge dx^3 + i \,  dx^1 \wedge dy^2 \wedge dy^3 \nonumber \\
&& -  \, dy^1 \wedge dy^2 \wedge dy^3. 
\eea
Note that these are the fluxes as in eq. (4.1) and (4.2) of \cite{Gomis:2005wc} up to the extra flux $\Delta F_3$. 
In particular, \cite{Gomis:2005wc} shows that this choice of $H_3$-flux satisfies properties ii) and iii).

To investigate property i) we evaluate the superpotential $W = \int \Omega \wedge (F_3 - \tau H_3)$ as proportional to
\bea
W &\propto& (1 + \tau_1 \tau_2) (1 + \tau_3 \tau) + \Delta W, \\
\Delta W &=&  M \left(\tau_1 -i\right)  \left(\tau_3 - \frac{i}{M}\right). 
\eea
This superpotential satisfies
\bea
\partial_\tau W &=& \tau_3 (1 + \tau_1 \tau_2) , \\
\partial_{\tau_1} W &=& \tau_2 (1 + \tau_3 \tau) + M\left( \tau_3 -  \frac{i}{M}\right), \\
\partial_{\tau_2} W &=& \tau_1 (1 + \tau_3 \tau), \\
\partial_{\tau_3} W &=& \tau (1 + \tau_1 \tau_2) + M (\tau_1-i).
\eea
The associated $F$-term conditions therefore yield a SUSY Minkowski vacuum ($W =0 = \partial_\tau W = \partial_{\tau_i} W$ ) with
all moduli $\tau_i$ and the axio-dilaton  stabilised at values
\bea
\tau_3 = \frac{i}{M}, \quad \tau_1 = i, \quad \tau_2 = i, \quad\ \tau = i M.
\eea
Note that $M>1$ leads to $g_s <1$ as required in the perturbative regime.

\bibliography{rev-inflation}  
\bibliographystyle{utphys}
\end{document}